\documentclass[
aps,
prl,
preprint,
floatfix,
superscriptaddress,
showkeys,
amsmath,
amssymb,
longbibliography,
floatfix]{revtex4-2}
\usepackage[english]{babel}
\usepackage{amsmath,amssymb,bbm,mathrsfs,bm,braket,color,graphicx,comment,amsfonts,dsfont}
\usepackage[colorlinks,linkcolor=blue,citecolor=blue,urlcolor=blue]{hyperref}
\usepackage[mathscr]{euscript}
\usepackage{physics}
\usepackage{xcolor}
\usepackage{ulem}
\usepackage{bm}
\usepackage{orcidlink}
\usepackage{multirow}

\def\t#1{\textrm{#1}}

\def\beq{\begin{equation}}
\def\eeq{\end{equation}}

\begin{document}

\title{Intrinsic negative magnetoresistance from the chiral anomaly of multifold fermions}

\author{F. Balduini} \thanks{These two authors contributed equally}  \affiliation{IBM Research - Zurich, 8803 Ruschlikon, Switzerland}
\author{A. Molinari} \thanks{These two authors contributed equally} \affiliation{IBM Research - Zurich, 8803 Ruschlikon, Switzerland}
\author{L. Rocchino}\affiliation{IBM Research - Zurich, 8803 Ruschlikon, Switzerland}
\author{V. Hasse}\affiliation{ Max Planck Institute for Chemical Physics of Solids, 01187 Dresden, Germany}
\author{C. Felser }\affiliation{ Max Planck Institute for Chemical Physics of Solids, 01187 Dresden, Germany}
\author{M. Sousa}\affiliation{IBM Research - Zurich, 8803 Ruschlikon, Switzerland}
\author{C. Zota}\affiliation{IBM Research - Zurich, 8803 Ruschlikon, Switzerland}
\author{H. Schmid}\affiliation{IBM Research - Zurich, 8803 Ruschlikon, Switzerland}
\author{A. G. Grushin} \email{adolfo.grushin@neel.cnrs.fr} \affiliation{Univ. Grenoble Alpes, CNRS, Grenoble INP, 
Institut Néel, 38000 Grenoble, France}
\author{B. Gotsmann} \email{bgo@zurich.ibm.com} \affiliation{IBM Research - Zurich, 8803 Ruschlikon, Switzerland}

\date{\today}
\maketitle

\textbf{
The chiral anomaly, a hallmark of chiral spin-1/2 Weyl fermions, is an imbalance between left- and right-moving particles that underpins both high and low energy phenomena, including particle decay and negative longitudinal magnetoresistance in Weyl semimetals.
The discovery that chiral crystals can host higher-spin generalizations of Weyl quasiparticles without high-energy counterparts, known as multifold fermions, raises the fundamental question of whether the chiral anomaly is a more general phenomenon.
Answering this question requires materials with chiral quasiparticles within a sizable energy window around the Fermi level, that are unaffected by trivial extrinsic effects such as current jetting.
Here we report the chiral anomaly of multifold fermions in CoSi, which features multifold bands within $\sim0.85$ eV around the Fermi level.
By excluding current jetting through the squeezing test, we measure an intrinsic, longitudinal negative magnetoresistance.
We develop the semiclassical theory of magnetotransport of multifold fermions that shows that the negative magnetoresistance originates in their chiral anomaly, despite a sizable and detrimental orbital magnetic moment contribution, previously unaccounted for.
A concomitant non-linear Hall effect supports the multifold-fermion origin of magnetotransport.
Our work confirms the chiral anomaly of higher-spin generalizations of Weyl fermions, currently inaccessible outside the solid-state.
}

The spin-statistics theorem forces elemental fermions to have half-integer spin. 
For instance, the Hamiltonian of a three-dimensional spin-1/2 massless Weyl fermion is $H=\eta\hbar v\mathbf{k}\cdot \boldsymbol{\sigma}$, which is linear in momentum $\mathbf{k}$, has a characteristic velocity $v$, and is written in terms of Pauli matrices $ \boldsymbol{\sigma}$ that encode its spin-1/2 degree of freedom.
The parameter $\eta=\pm1$ defines the chirality of the Weyl fermion, that act as sources and sinks, i.e. monopoles, of Berry curvature, with a sign determined by $\eta$~\cite{Armitage18}.
Intriguingly, some structurally chiral materials can realize, as low energy quasiparticles, higher-spin generalizations of Weyl fermions~\cite{manes_existence_2012,Bradlyn2016,tang_multiple_2017,Chang:2018bb}.
Referred to as multifold fermions, they are governed by a Hamiltonian  $H=\eta \hbar v\mathbf{k}\cdot \mathbf{S}$ where $\mathbf{S}$ represents an effective spin degree of freedom.
Because $\mathbf{S}$ can represent matrices of any spin, including integer spin, these massless fermions cannot exist as elementary particles.
Nonetheless, multifold fermions also have a definite chirality and act as sources or sinks of Berry curvature. 
For example, the central, middle and top bands of a spin-1 fermion have associated monopole charges $C=-2,0,2$ contrary to Weyl fermions, whose bands are Berry monopoles of charge $C=\pm1$.

Chiral massless particles are distinguished from non-chiral particles by their response to external magnetic fields.
Applying an electric $\mathbf{E}$ and magnetic field $\mathbf{B}$ such that $\mathbf{E}\cdot \mathbf{B}\neq 0$, changes the relative density of positive and negative chirality quasiparticles.
This results in a finite chiral current, not expected on classical grounds where positive and negative chirality quasiparticle densities are equal~\cite{bertlmann2000anomalies}.
This quantum effect is referred to as the chiral anomaly.
The chiral anomaly of Weyl fermions is by now text-book material, first discovered as a contribution to pion decay~\cite{bertlmann2000anomalies}.

In contrast, because multifold fermions do not exist as elementary particles, their anomalies are much less studied theoretically~\cite{Ezawa2017,Lepori2018,Nandy2019}, and have no experimental confirmation.
Experimentally, one challenge lies in identifying a chiral semimetal that realizes chiral massless quasiparticles within a large window of energy around the Fermi energy. 
Even more dramatically, the main experimental consequence of the chiral anomaly, a longitudinal negative magnetoresistance~\cite{son_chiral_2013}, is often masked by a trivial effect known as current jetting~\cite{Arnold2016,ong_experimental_2021}, arising from an enhanced anisotropic conductivity in a magnetic field in metals.

\begin{figure*}[htbp]
\includegraphics[scale = 1.25]{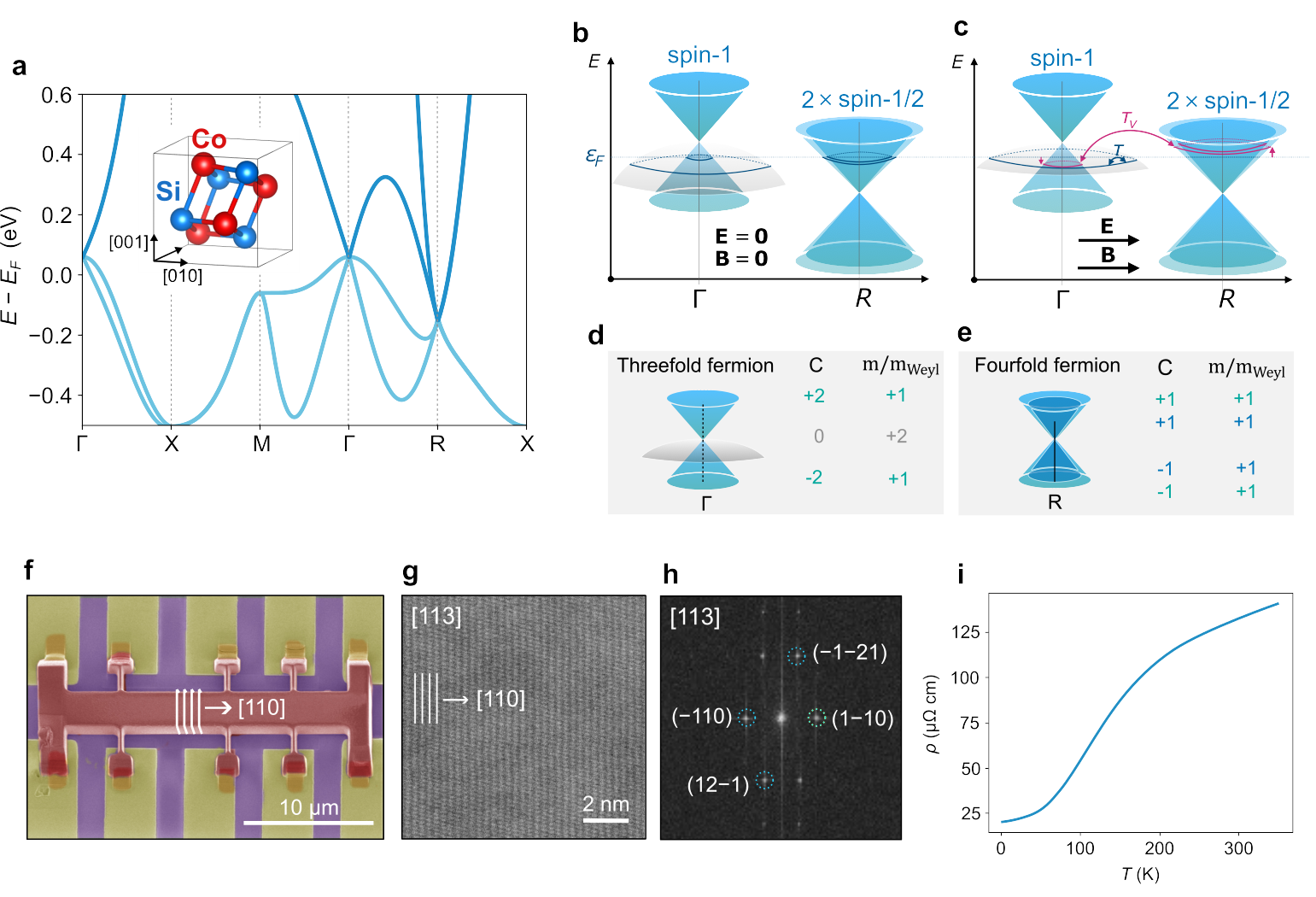}
\caption{\label{fig:wide1}{\textbf{Multifold fermions in CoSi and single-crystal sample.} 
{\bf{a}} Tight-binding band structure of CoSi crystal (inset) without spin-orbit coupling. The spin-1, threefold fermion and the double-spin-1/2 Weyl are located around $\Gamma$ and $R$, respectively. 
{\bf{b}} The linear and quadratic bands of the spin-1 threefold fermion at $\Gamma$ and the double spin-1/2 Weyl around $R$ at equilibrium. {\bf{c}} The chiral anomaly unbalances the density of spin-1 and double-spin-1/2 multifold fermions when $\bf{E} \cdot \bf{B}\neq 0$ through an internode scattering time $\tau_v$, resulting in a negative magnetoresistance. Intranode scattering (with characteristic time $\tau$), and a finite orbital magnetic moment contribute against the chiral-anomaly-induced negative magnetoresistance. 
{\bf{d}} The linear bands of the threefold fermion at $\Gamma$ have monopole charge of $\pm$2 and orbital magnetization equal to a Weyl fermion ($+1$). The parabolic band at $\Gamma$ has no Chern number and double orbital magnetization compared to a Weyl fermion ($+2$). 
{\bf{e}} The double spin-1/2 multifold fermion at $R$ is composed of two Weyl fermions of equal chirality, separated for clarity. Each has a monopole charge of $\pm$1 and orbital magnetization $+1$.
\textbf{f} False-colored scanning electron micrograph of our sample. The multi-terminal Hall bar was cut from a bulk single crystal using focused ion beam sculpturing.
\textbf{{g}} Scanning transmission electron microscopy (STEM) shows the atomically resolved structure of the crystal as seen perpendicular to the transport direction, indicating a [110] transport direction. 
\textbf{h} Diffraction pattern extracted by calculating the two-dimensional Fourier transform of the STEM image.
\textbf{i} Resistivity versus temperature of the device in \textbf{e}.
} }
\end{figure*}

Nevertheless, the chiral anomaly is theoretically expected for multifold fermions~\cite{Ezawa2017,Lepori2018}.  
In the ultraquantum limit of high-magnetic fields~\cite{Bradlyn2016,Ezawa2017}, a number of chiral Landau levels equal to the monopole charge of the band can pump left-moving to right-moving chiral fermions, just as in the case of Weyl fermions.
However, the ultraquantum limit remains inaccessible in real multifold materials.
For small magnetic fields, a semiclassical transport theory, along the lines of derivations for Weyl fermions, predicts a negative magnetoresistance rooted in the Berry curvature~\cite{Nandy2019} (see Fig.~\ref{fig:wide1}\textbf{b,c}).
However, existing semiclassical derivations neglect the orbital magnetic moment of multifold fermions~\cite{Flicker2018}, the self-rotating motion of a wave packet around the magnetic field~\cite{Xiao2010}, whose effect can be as sizable as the Berry curvature~\cite{Morimoto2016}.
It is thus unknown if their combined effect allows negative magnetoresistance arising from the chiral anomaly in multifold fermions to be observable.

Here we report the observation of the chiral anomaly in multifold fermions by measuring magnetotransport in Cobalt monosilicide (CoSi) and comparing it to a semiclassical theory that includes the Berry curvature and orbital magnetic moment.
We find that the orbital moment decreases, but not overcomes, the chiral anomaly contribution.
In agreement with our theoretical calculations, our measurements reveal a positive longitudinal magnetoconductance up to $1\%$ between $[-2.5,2.5]$ Tesla.
Its angular dependence is well described by the expected, but rarely observed $\cos^2\theta$ dependence, where $\theta$ is the relative angle between electric and magnetic fields.
These effects occur concomitantly with a non-linear contribution to the Hall effect, originating in both the Berry curvature and orbital magnetic moment, further supporting the multifold-fermion origin of the positive magneto-conductance.
We confirm that these properties are intrinsic to the material, as we can discard current-jetting effects using a recently proposed squeezing test~\cite{Liang:2018cj}. 

CoSi is an example of a chiral crystal with multifold fermions at the Fermi level, Fig.~\ref{fig:wide1}.
It belongs to a family of chiral crystals whose crystal symmetries enforce multifold fermions as low energy quasiparticles. 
The cleanest experimentally confirmed multifold materials are the chiral crystals AlPt and the monosilicides CoSi, RhSi, all in space-group 198~\cite{takane_observation_2019,raoNature2019,sanchez_topological_2019,Schroeter2019,Schroter179,Yao:2020cc,Sessi:2020dw}. 
Among them, CoSi is remarkably simple.
Ignoring spin-orbit coupling, which is relatively weak compared to RhSi and AlPt, CoSi has been confirmed to display multifold fermions within a large window, of order $0.85$ eV around the Fermi level (see Fig.~\ref{fig:wide1}\textbf{a}).
The band structure obtained from a symmetry-compatible tight-binding model of this material~\cite{chang_unconventional_2017,Flicker2018}, fitted to ab-initio calculations~\cite{Xu:2020ei,Ni:2021cta}, shows a three-band spin-1 fermion at the Brillouin center (Fig.~\ref{fig:wide1}\textbf{d}), and a four-fold crossing at the Brillouin corner R point composed by two copies of a spin-1/2 Weyl fermion of equal chirality that meet in a single point (Fig.~\ref{fig:wide1}\textbf{e}).
This band-structure explains well features seen both in photoemission and optical experiments~\cite{sanchez_topological_2019,takane_observation_2019,raoNature2019,Xu:2020ei,Ni:2021cta,Flicker2018}.

For our transport experiments, we grew CoSi single crystals using the chemical vapor transport method (see Supplementary Information). 
Using focused ion beam we fabricated a micro-Hall bar starting from a lamella extracted from the CoSi single crystal,  Fig.~\ref{fig:wide1}\textbf{f}. 
The microfabrication allows for good control of geometry and crystalline direction, and an even distribution of the magnetic field over the sample.
To ensure accurate alignment of the sample, we extracted a second lamella from the same single crystal and oriented it in the same direction as the one used for the Hall bar fabrication. 
The second lamella underwent scanning transmission electron microscope (STEM) analysis to verify the alignment of the crystalline axes (Fig.~\ref{fig:wide1}\textbf{g-h}). 

In Fig.~\ref{fig:wide1}\textbf{i}, we show the resistivity as a function of temperature. 
We observe that at low temperatures, the resistivity is $\rho_{xx}(2K) = 20$ $\mu\mathrm{\Omega cm}$, and monotonically increases with temperature ($RRR = \rho (300\,{\rm{K}}) / \rho (2\,{\rm{K}})  \approx 6$), which aligns with the literature values from bulk crystals \cite{xu_crystal_2019}. 
Typically, the chiral anomaly is observed in Weyl semimetals with semiconducting-like resistivity \cite{hirschberger_chiral_2016, niemann_chiral_2017, xiong_evidence_2015, li_negative_2016, yang_chiral_2015}, indicating the proximity of the Fermi level to the Weyl points.
However, this is not the case for CoSi. 
Due to its large topologically-non trivial energy window, it becomes feasible to observe multifold-related effects even at relatively high carrier densities. 
This behavior sets CoSi apart from traditional Weyl semimetal systems.
\begin{figure*}[htbp]
\includegraphics[scale=1]{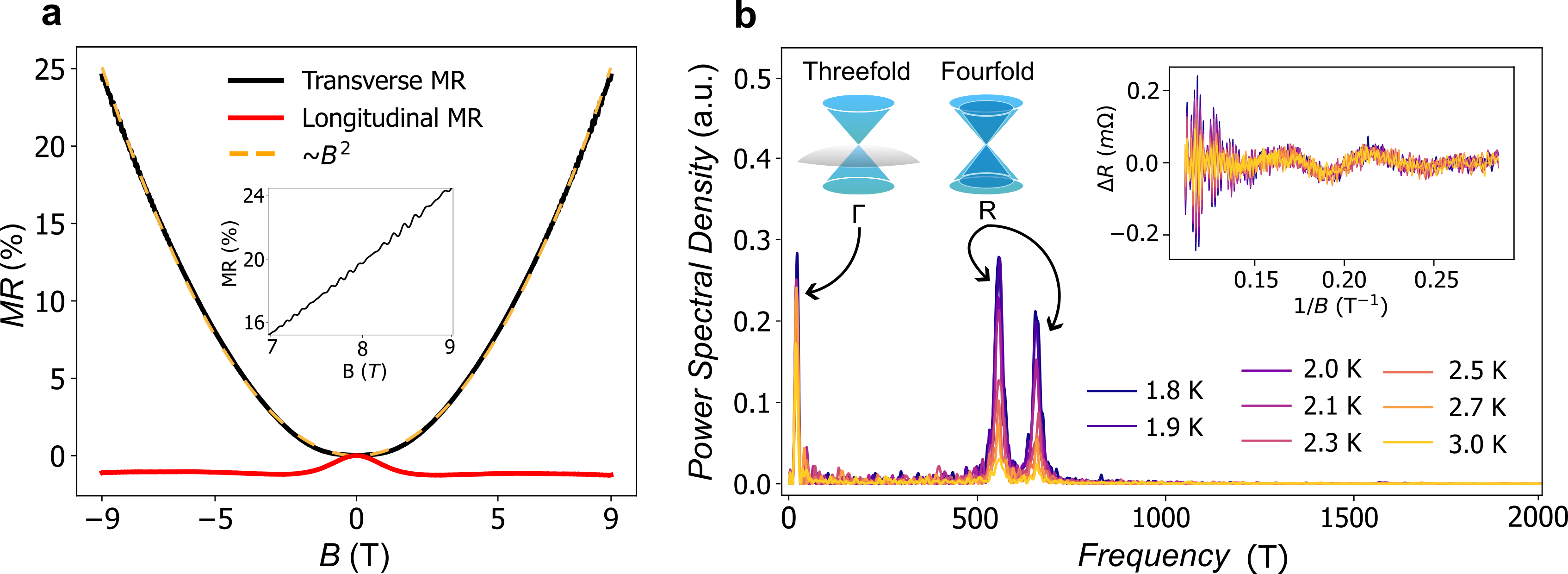}
\caption{\label{fig:MR}{\textbf{Magnetoresistance of CoSi.} \textbf{a} Magnetoresistance ($MR = (R(B=0) - R(B))/R(B=0)$) in transversal geometry ($B \perp I$, black line) and longitudonal geometry ($B || I$, red line), at 2 K. The approximate parabolic shape of the transversal MR (TMR) is demonstrated using a parabolic fit (dashed yellow line). The inset shows Shubnikov-de\,Hass oscillations in the TMR.
\textbf{b} Analysis of the Shubnikov-de\,Hass oscillations. The residual resistance change $\Delta R$ after subtracting the background plotted versus the inverse of applied magnetic field $B$ shows several frequency contributions (see inset). The power spectral density of $\Delta R$ in arbitrary units shows three clear frequencies for temperatures between 1.8 and 3\,K. These can be assigned to the three and fourfold fermions at $\Gamma$ and $R$, respectively, confirming that both fermions contribute to transport.}} 
\end{figure*}

To further characterize the carrier types and electronic pockets at the Fermi level, we now apply a magnetic field perpendicular to the current direction.
We observe the resistivity increases quadratically up to 24$\%$ at 9\,T and 2\,K, with no sign of saturation (Fig.~\ref{fig:MR}\textbf{a}). 
Shubnikov-de Haas (SdH) oscillations are present at frequencies of 21 T, 556 T, and 657 T (Fig.~\ref{fig:MR}\textbf{b}). Following previous studies \cite{wang_haas--van_2020, xu_crystal_2019, guo_hidden_2021,  sasmal_shubnikov-haas_2022}, we assign these frequencies to the tiny hole pocket at $\Gamma$ and the double-Weyl electron pockets at $R$, respectively, indicating the presence of multifold fermions in our sample. 
The Onsager relation in combination with the Luttinger theorem allows extracting the carrier density in $\Gamma$: $n_{\Gamma} = 5.4 \cdot 10^{17}$  $\mathrm{cm^{-3}}$ and in $R$: $n_{R} = 1.7 \cdot 10^{20}$  $\mathrm{cm^{-3}}$, in excellent agreement with the value found from the Hall effect at low temperatures $n_{H} = 1.7 \cdot 10^{20}$  $\mathrm{cm^{-3}}$, which reveals an electron dominated transport (Supplementary Information). 
\begin{figure*}[htbp]
\includegraphics[width=\textwidth]{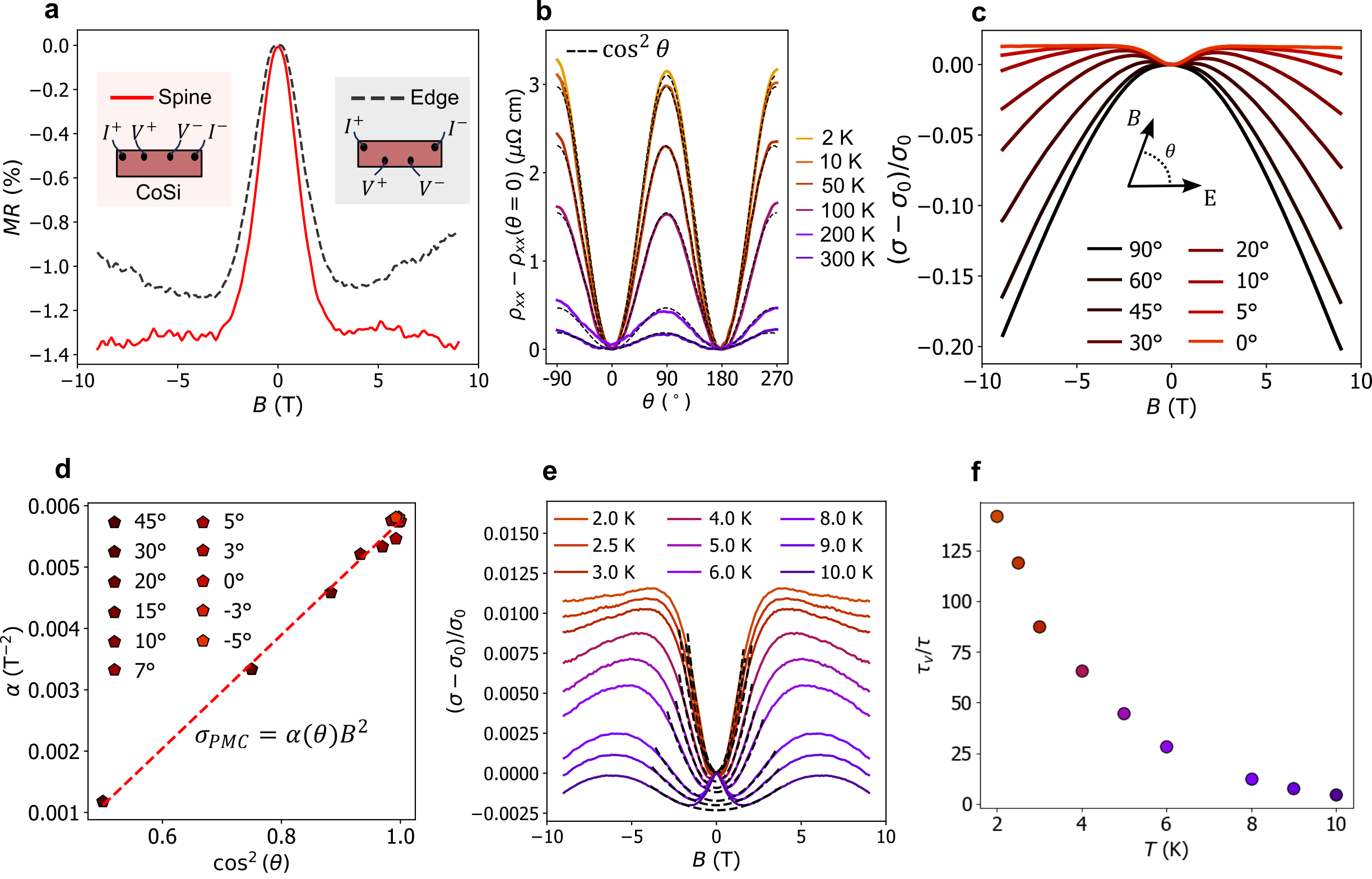}
\caption{\label{fig:chiral}{\textbf{Negative longitudinal magnetoresistance from the chiral anomaly of multifold fermions}. \textbf{a} Result of the squeeze test. The small variation of the longitudinal magnetoresistance upon changing the probing contact geometries demonstrates that the measured phenomenon is intrinsic to the material. \textbf{b} Dependence of the resistivity at an applied magnetic field of $B=9$\,T on the angle $\theta = \angle (B,E)$ at various temperatures. Dashed lines are fits to the expected cos$^2(\theta)$ dependence. \textbf{c}
 Normalized conductance variation measured at 2 K, as a function of the applied magnetic field for some angles $\theta$ between $B$ and $E$.
\textbf{d} At small angles $\theta$, the magnetoconductivity increases quadratically with magnetic field $(\sigma-\sigma_0)/\sigma_0 = \alpha B^2$. The parabolic coefficient $\alpha$ increases as $\cos^2(\theta)$, as expected for the chiral anomaly. \textbf{e} Longitudinal magnetoconductance at various temperatures together with parabolic fits (black dashed lines).
\textbf{f} Internode scattering time $\tau_v$ over intranode scattering time $\tau$ as function of temperature extracted using the fits in \textbf{e} and Eq.~\eqref{eq:allmain}. }}
\end{figure*}

Having characterized our sample, Fig.~\ref{fig:MR}\textbf{a} shows the longitudinal magnetoresistance (MR). 
When the magnetic field is rotated in the direction parallel to the electrical current, the MR exhibits a negative trend, decreasing as ${B^2}$ until it saturates at approximately 2.5 T. 
It is worth noting that previous research reported a positive longitudinal magnetoresistance in CoSi single crystal \cite{petrova_magnetoresistance_2023}, yet asymmetric with the magnetic field. In contrast, a negative longitudinal magnetoresistance in CoSi has been reported in \cite{guo_hidden_2021} and in Fe-doped CoSi in \cite{schnatmann_signatures_2020}.

The chiral anomaly is expected to leave an imprint as a negative longitudinal magnetoresistence~\cite{son_chiral_2013}. 
A negative magnetoresistance has been observed in the Dirac semimetals $\mathrm{Na_3Bi}$ \cite{xiong_evidence_2015}, $\mathrm{Cd_3As_2}$ \cite{li_negative_2016}; in the type I Weyl semimetal TaAs \cite{zhang_signatures_2016, huang_observation_2015}, NbAs \cite{yang_chiral_2015} and NbP \cite{niemann_chiral_2017}; in the type II Weyl semimetal $\mathrm{WTe_2}$ \cite{wang_gate-tunable_2016}; in the heavy fermion semimetal YbPtBi \cite{guo_evidence_2018}; in GdPtBi \cite{hirschberger_chiral_2016} and $\mathrm{ZrTe_5}$ \cite{liang_anomalous_2018} where Zeeman energy leads to band crossings and the formation of Weyl nodes. 
In all of these materials, Weyl nodes are fairly close in momentum space and constrained to a small energy window ($\approx 50-100$ meV). 
In comparison, CoSi is advantageous because multifold fermions of opposite chiralities are maximally separated in the Brillouin zone and exist in a large energy window ($\approx 0.85$ eV).

However, before analyzing the longitudinal magnetoresistance it is necessary to confirm that it is intrinsic.
Notably, a longitudinal magnetic field that enhances the anisotropy of the conductivity tensor is known to lead to a spurious negative or positive magnetoresistance depending on the location of the contacts~\cite{pippard_magnetoresistance_nodate,Arnold2016,ong_experimental_2021}.
To exclude this phenomenon, known as current jetting, we performed the squeeze test proposed in Ref.~\cite{ong_experimental_2021}.
The squeeze test compares different current inlet and voltage probe geometries of the CoSi microbar. 
Fig.~\ref{fig:chiral}\textbf{a} shows the most divergent results. The measurements consistently show a negative longitudinal magnetoresistance, regardless of the chosen contact configuration, providing evidence of negligible current jetting \cite{ong_experimental_2021}. 
The absence of current jetting phenomena is also consistent with the relatively modest electron mobility of CoSi extracted from the Hall effect $\mu = 3 \cdot 10^{3}$ $\mathrm{cm^{2}/Vs}$ (see Supplementary Information).

We note that other phenomena, such as localization effects, mobility fluctuations, magnetism, or finite-size effects, could potentially contribute to a negative longitudinal magnetoresistance. 
However, we have ruled out these possibilities due to the smooth parabolic behavior observed in both transverse and longitudinal MR at low temperatures, yet with opposite sign, contrary to what is expected in the case of localization effects \cite{naumann_orbital_2020}, disorder \cite{breunig_gigantic_2017} or mobility fluctuations \cite{schumann_negative_2017}. The consistently monotonic trend of the negative MR of CoSi at 2 K indicates that finite-size effects do not underlie its origin \cite{pippard_magnetoresistance_nodate}. Finally, while an excess of Cobalt could contribute to the negative MR, because of its magnetic properties, such an occurrence would typically coincide with hysteresis in the MR or Hall effect data \cite{molinari_disorder-induced_2023}, which is not evident within our experimental error (see Supplementary Information).
 In Fig.~\ref{fig:chiral}\textbf{b} we show the change in resistivity as a function of angle $\theta$ between the current and the magnetic field, fixed at 9 T. The experimental data show a clear $\cos^2 \theta$ dependence, indicated by the dashed black lines.
 We observe that, when the magnetic field is aligned with the current, the positive magnetoconductance phenomenon is maximal (Fig.~\ref{fig:chiral}\textbf{c}). The positive magnetoconductance increases as $\cos^2 \theta$ for $\theta$ approaching 0$^{\circ}$, as demonstrated in Fig.~\ref{fig:chiral}\textbf{d}.

While the $\cos^2\theta$ is expected for the negative magnetoresistance-induced chiral anomaly, it contrasts with what is observed in Weyl materials, which show a strong narrowing ($\cos^n\theta$ with $n>2$)~\cite{xiong_evidence_2015,zhang_signatures_2016,ong_experimental_2021}, which has been attributed to anisotropy in disorder~\cite{Behrends2017} or magneto-transport~\cite{Deng2020}. To our knowledge, the only exception is the heavy-fermion semimetal YbPtBi~\cite{guo_evidence_2018}.
Narrowing effects seem to be absent or weak in our samples, and are consistent with the absence of spurious effects, further supporting that we are measuring an intrinsic contribution.

To interpret the magnetoresistance data we have developed a semiclassical theory for magnetoresistance~\cite{son_chiral_2013} to the case of multifold fermions.
The main difference compared to previous semiclassical work is the inclusion of the orbital magnetic moment.
This may seem a nuanced point but, as noticed in Refs.~\cite{Morimoto2016,Agarwal21} for the case of spin-1/2 Weyl fermions, the orbital magnetic moment contributes with a magnitude comparable to the Berry curvature to the magnetoresistance of Weyl fermions.
Hence, in the multifold case, it can cause a sizable under- or over-estimation of the effect, or even completely suppress negative magnetoresistance.

Including all contributions, we find that the magnetoresistance is dominated by the spin-1 threefold linear hole band at $\Gamma$ and the double-spin-1/2 electron pocket at $R$, given by (see Supplementary Information):
\begin{equation}
\frac{\sigma-\sigma_0}{\sigma_0} = \dfrac{\left(A+B\frac{\tau_v}{\tau}\right)\cos^2\theta-C}{20\left(1+2r_2 \right)}\dfrac{B^2}{\left(k^\Gamma_F\right)^4}\dfrac{e^2}{\hbar^2},
\label{eq:allmain}
\end{equation}

where $A=-9+2 r_1$, $B=10\left(3+r_1\right)$, $C=\left(2+4r_1\right)$, $r_1=\left(\frac{k^{\Gamma}_F}{k^{R}_F}\right)^2\frac{v_{R}}{v_{\Gamma}}$
and $r_2=\left(\frac{k^{R}_F}{k^{\Gamma}_F}\right)^2\frac{v_{R}}{v_{\Gamma}}$.
If the ratio between inter ($\tau_v$) and intranode ($\tau$) scattering times is $\frac{\tau_v}{\tau}>1$ this implies $\frac{\sigma-\sigma_0}{\sigma}>0$ for any $B$.
At $\theta=0$ the magnetoconductance increases quadratically with the magnetic field, as seen in our samples, Fig.~\ref{fig:chiral}\textbf{e}.
The calculated angle-dependent conductivity, proportional to $\cos^2\theta$, is also in agreement with our experiment, Fig.~\ref{fig:chiral}\textbf{d}.
By tracking the orbital magnetic moment contributions to the conductivity, we observe that these work against, but crucially do not overcome, the positive chiral anomaly terms. 
Hence, the observation of intrinsic positive magnetoconductance is a signature of the chiral anomaly of multifold fermions, even when we take into account the large orbital magnetic moment of the multifold fermions.

To be more quantitative, we can make use of the fact that $k^{\Gamma}_F = 0.25$ nm$^{-1}$ $\approx 5.2 k^{R}_F$, which we extract from our quantum oscillation measurements. 
The tight-binding model fitted to ab-initio calculations shows that $v_{\Gamma}=v_{R}/\sqrt{3}$~\cite{Xu:2020ei,Ni:2021cta}.
Using these numbers we can extract the ratio $\tau_v/\tau$ by fitting $(\sigma-\sigma_0)/\sigma_0$ to our measurement. The fit is shown in Fig.~\ref{fig:chiral}\textbf{e}, which leads to $\tau_v/\tau\approx10^2$ at low temperatures (Fig.~\ref{fig:chiral}\textbf{f}). The large ratio $\tau_v/\tau$ is indicative of a long-lived chiral current, as expected for multifold fermions with maximally separated chiral branches.

\begin{figure*}[htbp]
\includegraphics[width=\textwidth]{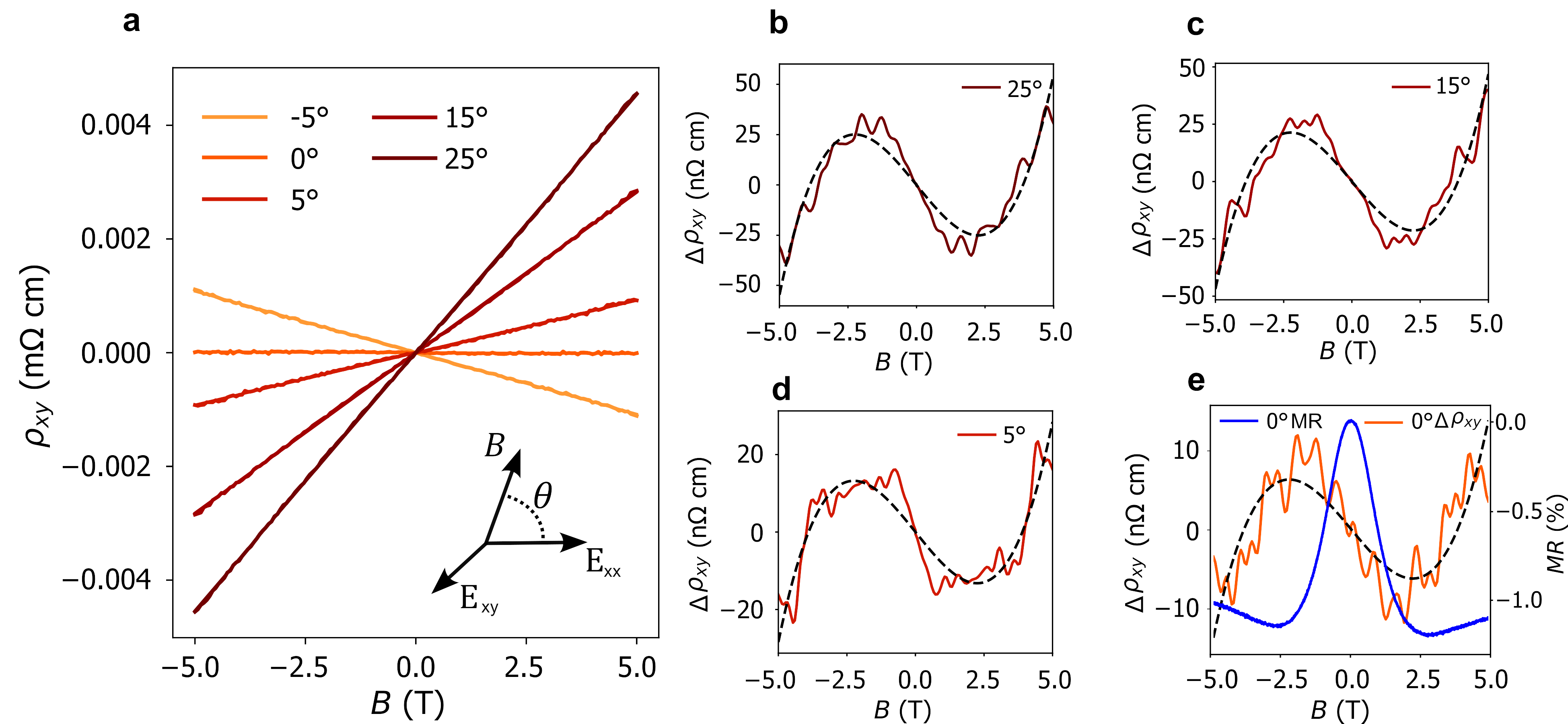}
\caption{\label{fig:ahe}{\textbf{Cubic-in-$B$ Hall effect.} \textbf{a} Angle-dependent Hall effect. \textbf{b-d} A non-linear-in-$B$ Hall effect is visible after subtracting the linear component from the Hall effect. The remaining contribution is well described by cubic polynomial $\sim \alpha B + \beta B^3$, as indicated by the dashed black lines. 
The negative longitudinal magnetoresistance is plotted on top \textbf{e}, showing that the field range where the cubic component of the Hall effect is non-zero coincides with the range of decreasing negative magnetoresistance.}}
\end{figure*}

To further validate the multifold origin of the observed negative longitudinal magnetoresistance, we conducted angle-dependent Hall effect measurements to search for indications of a topological Hall effect, which would be indicative of non-zero Berry curvature and non-trivial material topology, given the non-magnetic nature of CoSi. After subtracting a linear fit from the Hall data shown in Fig.~\ref{fig:ahe}\textbf{a}, we observe a cubic-in-$B$ contribution to the Hall effect, as depicted in Fig.~\ref{fig:ahe}\textbf{b-e}. Remarkably, the observed trends closely resemble those reported in YbPtBi \cite{guo_evidence_2018} and ZrTe$_5$ \cite{liang_anomalous_2018}, reinforcing the idea of a topological connection. We observe that the range of the magnetic field where the cubic-in-B Hall effect varies coincides with the range where the longitudinal magnetoresistance decreases quadratically before saturation (Fig.~\ref{fig:ahe}\textbf{e}). In the Supplementary Information, we show that the same semiclassical theory that leads to \eqref{eq:allmain} predicts that the leading contributions to the cubic-in-$B$ Hall effect are due to the orbital magnetic moment and the Berry curvature of the multifold nodes. 
Since both diverge close to the multifold nodes, a non-linear cubic Hall is expected to be sizable where the positive magnetoconductance is sizable, as seen in our experiment.
Concretely, we find that the dominant contribution is the filled linear band at $\Gamma$, as its Fermi momentum is closest to the multifold crossing.

In summary, our measurements confirm the existence of negative longitudinal magnetoresistance in single-crystal CoSi, which we attribute to the chiral anomaly of multifold fermions.
Our theory shows that the orbital magnetic moment contributions work against the chiral anomaly terms, but do not overcome them, allowing negative magnetoresistance to be observable in our experiment. 
Additionally, the orbital and Berry curvature contributions dominate the cubic-in-magnetic field Hall term, acting as an additional signature of multifold fermions.
While high-energy physics phenomena have been previously observed in analog condensed matter systems, our results mark the observation of a quantum anomaly of particles with no counterpart in high-energy physics, as multifold fermions are forbidden to exist as elementary particles.
Our work showcases that quantum materials are a fruitful avenue to formulate and observe the most general physical phenomena linked to quantum anomalies.

\section{\label{sec:level1}Methods}

\subsection{\label{sec:level1}Crystal growth}
CoSi single crystals were grown in Te-flux. The starting materials Co (99.95\%, 20 Alfa Aesar), Si (99.999\%, Chempur) and Te (99.9999\%, Alfa Aesar) were mixed in the molar 21 ratio of 1:1:20 and heated to 1050° C at a rate of 100° C/h and held there for 15 h. Successively, the sample was cooled to 700° C at a rate of 2° C/h, and extra Te-flux was removed by centrifugation. High-quality CoSi single crystals in the mm-range resulted from this growth protocol.

\subsection{\label{sec:level1}Sample fabrication}
 From the CoSi bulk sample, a 2 $\mu$m$ \times$ 1.5 $\mu$m $\times$ 15 $\mu$m lamella was cut and then patterned to Hall bar shape with a focused ion beam system (FIB) of the type FEI Helios 600i using 30 keV Ga+ ions. The micro Hall bar was then transferred onto a patterned chip with Au contact pads (150 nm Au + 10 nm Ti, for adhesion) and weld using ion-assisted deposition of Pt.

\subsection{\label{sec:level1}Electrical transport measurements}
Electrical measurements were performed in a cryostat (Dynacool from Quantum Design) using external lock-in amplifiers (MFLI from Zurich Instruments). The electrical current is always applied along the [110] direction. B oriented in [113] or [110].

\subsection{\label{sec:level1}Scanning Transmission Electron Microscopy}
The STEM measurements have been performed with a double spherical aberration corrected JEOL ARM200F microscope operated at 200 kV.

\section{Acknowledgements}

AGG is grateful to J. H. Bardarson, J. Behrends, and D. Pesin for insightful discussions. AGG, CZ, AM and BG acknowledge financial support from the European Union Horizon 
2020 research and innovation program under grant agreement No. 829044 (SCHINES). F.B. and B.G. acknowledge the SNSF project HYDRONICS under the Sinergia grant (No.189924). A.M. acknowledges funding support from the European Union’s Horizon2020 research and innovation program under the Marie Sklodowska-Curie Grant Agreement No. 898113 (InNaTo).
We are grateful to Philip Moll for sharing insights and support in FIB-based microstricturing.
We wish to acknowledge the support of the Cleanroom
Operations Team of the Binning and Rohrer Nanotechnology
Center (BRNC).
Continuous support from Ilaria Zardo, Heike Riel, Mark Ritter, and Kristin Schmidt is gratefully acknowledged.

\clearpage
\newpage

\setcounter{secnumdepth}{5}
\renewcommand{\theparagraph}{\bf \thesubsubsection.\arabic{paragraph}}

\renewcommand{\thefigure}{S\arabic{figure}}
\setcounter{figure}{0} 

\appendix

\section{Extended Data}
\begin{figure*}[htbp]
\includegraphics[scale=1]{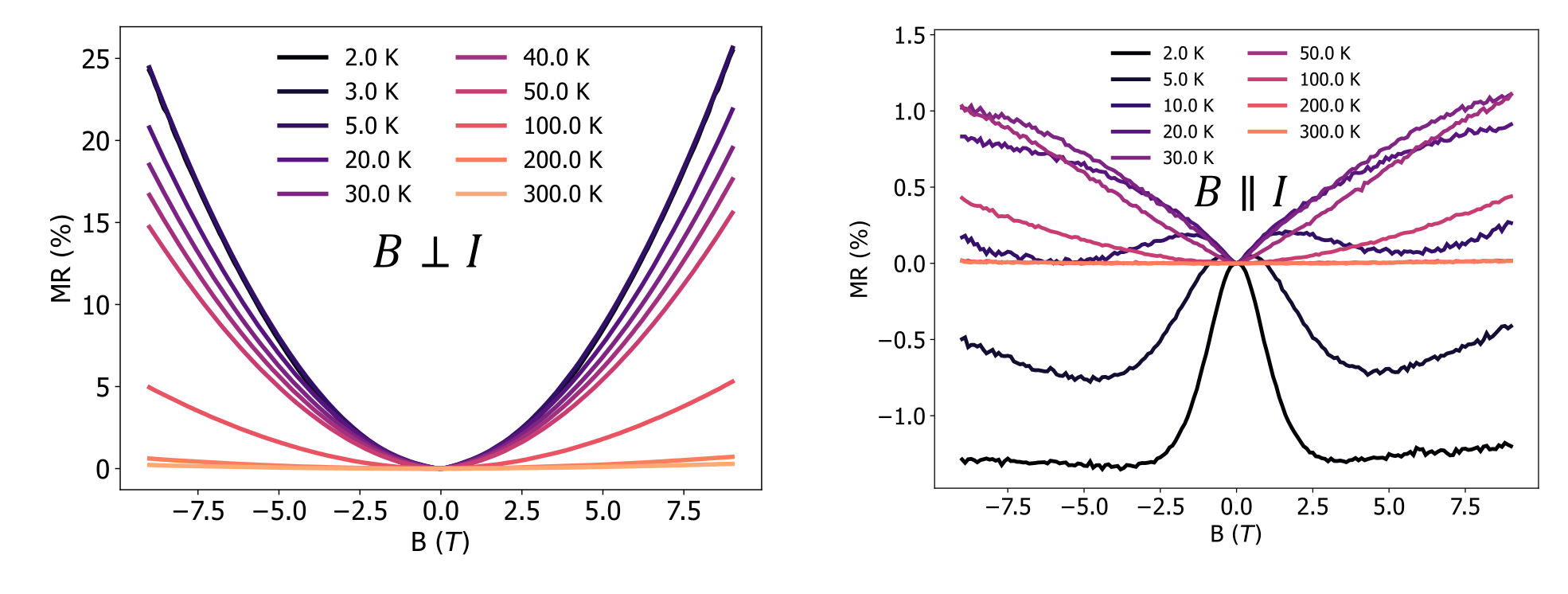}
\caption{{\textbf{Temperature dependence of the MR.}  Transverse MR $B \perp I$ (\textbf{left}) and longitudinal MR $B || I$ (\textbf{right}).}} 
\end{figure*}

\clearpage
\newpage

\begin{figure*}[htbp]
\includegraphics[scale=0.9]{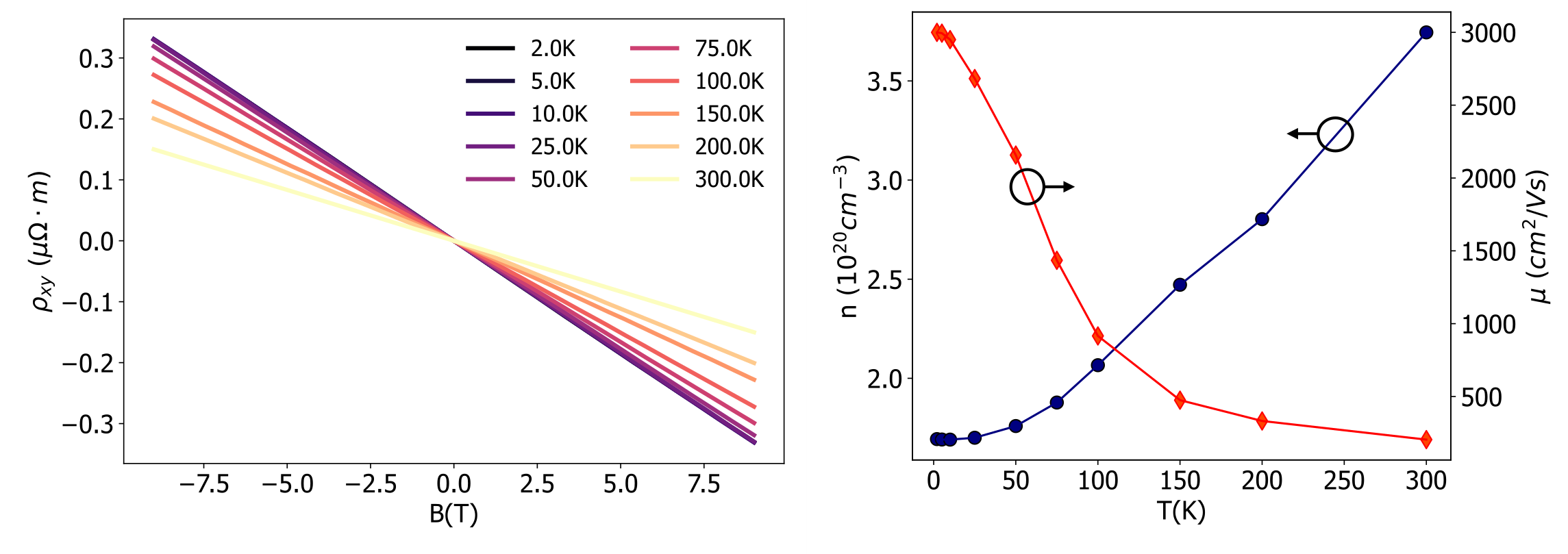}
\caption{{\textbf{Hall effect} Hall resistance (\textbf{left}) and carrier density and mobility extracted from a linear fit of the Hall effect $\rho_{xy}=\frac{B}{ne}$, where $n$ is the carrier density and $e$ the electron charge. The mobility is calculated as $\mu = (ne\rho_{xx})^{-1}$ (\textbf{right}). }} 
\end{figure*}

\clearpage
\newpage

\begin{figure*}[htbp]
\includegraphics[scale=1]{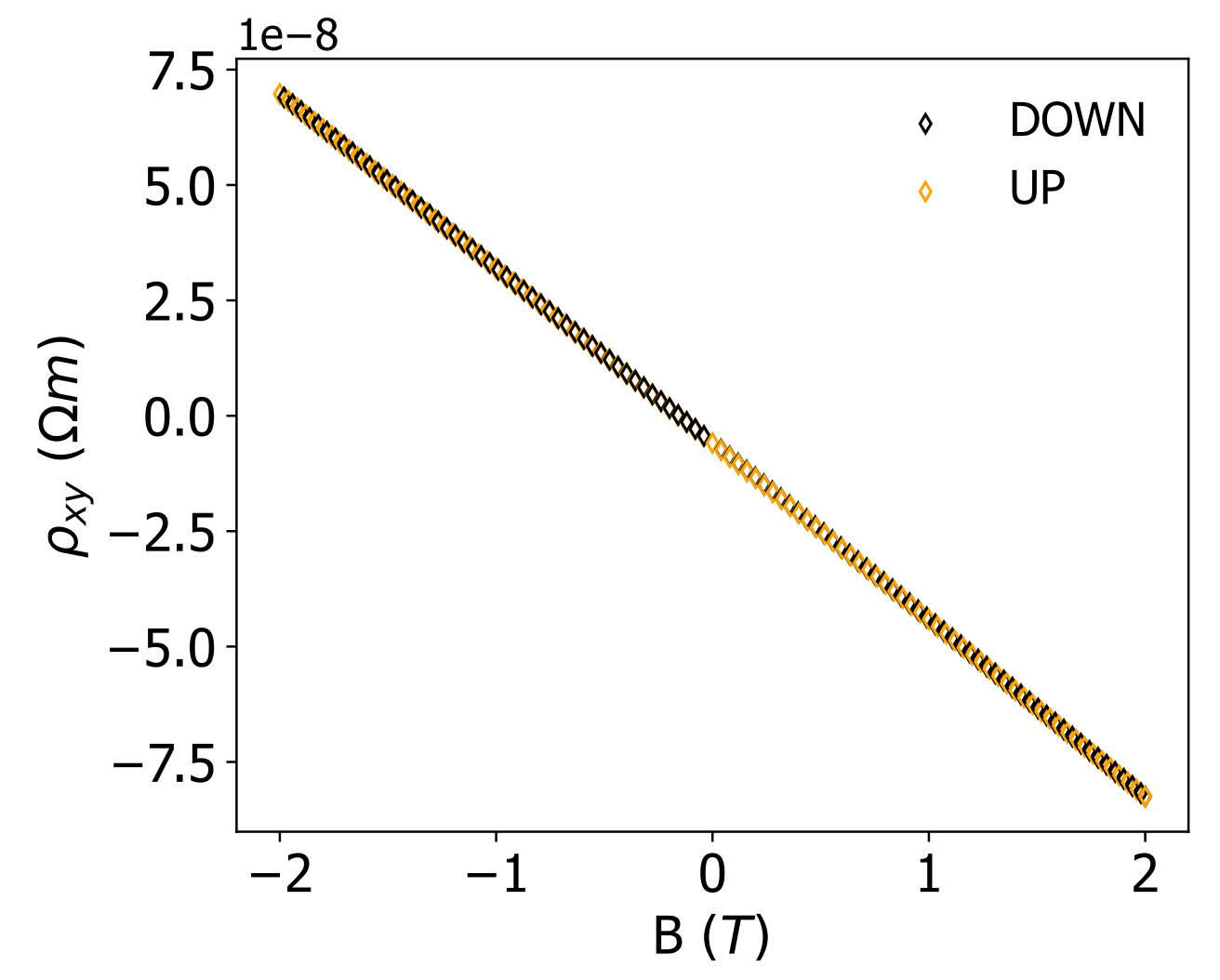}
\caption{{\textbf{Bidirectional Hall effect measurement.} The Hall signal does not show any hysteresis, supporting the hypothesis of negligible magnetic impurities in the sample.}} 
\end{figure*}

\clearpage
\newpage

\section{Introduction to multifold fermions in CoSi}
Higher-spin fermions have total monopole charges at the Fermi level of $1 < |C| \leq 4$ \cite{Bradlyn2016,Flicker2018}, generally surpassing Weyl nodes, which have $|C|=1$. 
As we review next, the chiral-anomaly-induced negative magnetoresistance is proporional to $C$, and thus it is larger 
for higher-spin fermions than for a single Weyl nodes, given the same Fermi velocity.

The Hamiltonian of a generic chiral multifold fermion of (pseudo)spin $S$  around a time-reversal invariant momentum, is, to lowest order in momentum $\mathbf{k}$, given by
\begin{equation}
\label{eq:multifoldlowenergy}
    H_\mathbf{k} = v \mathbf{k}\cdot \mathbf{S}(\alpha).
\end{equation}
The vector $\mathbf{S}$ is composed out of of three matrices that depend in general on band-structure parameters, $\alpha$, and do not form a spin-$S$ representation of SU(2).
However, neglecting spin-orbit coupling, as pertinent for CoSi, forces the matrices in  $\mathbf{S}(\alpha)$ to realize a representation of the spin algebra of spin-$s$.
Which may choose our convention such that, to linear order in momentum, the band structure at the $\Gamma$ point is given by $\epsilon_S = m_S v |\mathbf{k}| $ and $\epsilon_S = 2 m_S v |\mathbf{k}| $ for integer and half-integer spin, respectively, with $m_s=-S,-S+1,\cdots,S-1,S$, and $v$ the Fermi velocity.

At the $\Gamma$ point and without spin-orbit coupling the three-fold irreducible representation (irrep) of the little group of CoSi at the $\Gamma$ point enforces that $\mathbf{S}$ is a representation of $3\times3$ spin-1 matrices\cite{Bradlyn2016,Flicker2018}. 
Hence, to linear order in momentum, the band structure at the $\Gamma$ point is given by $\epsilon_\Gamma = m_1 v_\Gamma |\mathbf{k}| $ with $m_1=-1,0,+1$.
The Berry curvature of band $m_1$ is given by $\boldsymbol{\Omega}^{m_1}_\mathbf{k}=m_1\frac{\mathbf{k}}{|\mathbf{k}|^3}$,  corresponding to a charge $C=\pm 2$  monopole for bands $m_1 = \pm 1$ and charge $C=0$ monopole for band $m_1=0$.

At the corner of the Brillouin zone $R=(\pi,\pi,\pi)$ a four-fold irrep enforces a double spin-1/2 Weyl fermion with Hamiltonian
\begin{equation}
    H_\mathbf{k} = v_R \mathbf{k}\cdot \mathbf{\sigma}\otimes \tau_0,
\end{equation}
where $\tau_0$ encodes the Weyl copies at $R$. The band structure is given by two doubly-degenerate Weyl dispersion relation $\epsilon_R = \pm v_R |\mathbf{k}|$.
For each copy (there are four counting spin), the Berry curvature of band $\pm$ is given by $\boldsymbol{\Omega}^{\pm}_\mathbf{k}=\pm\frac{\mathbf{k}}{2|\mathbf{k}|^3}$, corresponding to charge $C=\pm 1$ monopoles.

The chiral anomaly of higher-spin fermions is proportional to its monopole charge $C$. 
A monopole charge of $C$ implies that the chemical potential will cross $C$ chiral Landau levels~\cite{Bradlyn2016,Ezawa2017}. 
Following the arguments by Nielsen and Ninomiya~\cite{NielNino81a,NielNino81b}, an electric field $\mathbf{E}$ with a component parallel to the direction of magnetic field $\mathbf{B}$ will increase or decrease the chemical potential of positively or negatively dispersing chiral Landau levels, respectively.
Both contributions add up and hence, the chiral anomaly equation for a single multifold fermion of charge $C$ becomes
\begin{equation}
\label{eq:CA}
    \partial_\mu j^{\mu} = |C| \left(c_A \mathbf{E}\cdot \mathbf{B}\right),
\end{equation}
where $j_{\mu}=(\rho,\mathbf{j})$ is the current four-vector, composed of the charge density $\rho$ and charge current density $\mathbf{j},$ and $c_A=\dfrac{1}{4\pi^2}$ is the chiral anomaly coefficient in units of $e^2/\hbar$. 
It has been noted that, when the chemical potential is exactly at the spin-1 crossing point, a linear model predicts that the chiral anomaly is equal to that of a single Weyl node~\cite{Lepori2018}. 
While this is an interesting regime, it is never achieved experimentally in CoSi.
Hence  in the following we calculate the chiral-anomaly-induced magneto-conductivity coming from the $R$ and $\Gamma$ multifold fermions semi-classically.

\section{Semiclassical Boltzmann transport}

This section contains the main theoretical results behind our interpretation of the longitudinal magnetoresistance. The three-fold and the double Weyl fermions, found in CoSi, induce negative magnetoresistance (positive magnetoconductance). As in other topological semimetals, this result is a consequence of their finite Berry curvature. Our calculations include the orbital magnetic moment, which acts to reduce, but not destroy the chiral anomaly induced negative magnetoresistance.

Our calculation generalizes standard derivations of positive magnetoconductance \cite{son_chiral_2013,Lundgren2014,KimKim2014,MaPesin2015,Nandy2019,Morimoto2016,Imran2018,Deng2019,Deng2020,Mandal2022} for Weyl fermions. The positive magnetoconductance of the double Weyl at the $R$ point follows from these works. The $R$ point is well approximated by two independent Weyl fermions of the same chirality.
We repeat this derivation here for pedagogical reasons, including the aforementioned orbital magnetic moment.

As for the $\Gamma$ point, the positive magneto-conductance can be derived in the ultraquantum, Landau level limit \cite{Ezawa2017}. However, CoSi is far from the ultraquantum regime, and our results need to be understood in the semiclassical regime of low magnetic fields.
Ref.~\cite{Nandy2019} derived the positive magneto-conductance of three-fold fermions. However, this reference did not include the orbital magnetic moment. We find that this quantity acts to decrease, but not overcome, the coefficient of the positive magnetoconductance found in Ref.~\cite{Nandy2019}. The orbital magnetic moment also contributes to a $B^3$ Hall contribution in the semiclassical regime. 

Our derivation below follows closely that of Refs.~\cite{Morimoto2016,Mandal2022} and uses the same notation. The main differences are that here we will be concerned only with DC response, we will expand the Hall response to cubic order in magnetic field, and we will calculate the responses of  threefold fermion and not only a Weyl fermion.

The motion of an electron wavepacket in a metal centred at position ${\bf r}$ and momentum ${\bf k}$ is governed by the semiclassical equations of motion~\cite{Sundaram99,Xiao2010}
\begin{subequations}
\begin{align}
\dot{\bm r}&=\frac{1}{\hbar}\bm{\nabla_k} \epsilon_{\bm k} - \dot{\bm k} \times \bm \Omega_\mathbf{k}, \\
\hbar \dot{\bm k}&= - e \bm E - e \dot{\bm r} \times \bm B.
\end{align}
\label{eq: EOM}
\end{subequations}
These equations include the electric and magnetic fields, $\bm E$ and $\bm B$ respectively, the Berry curvature 
\begin{align}
\bm \Omega_\mathbf{k}=- \t{Im}[\bra{\nabla_{\bm k} u_{\bm k}} \times \ket{\nabla_{\bm k} u_{\bm k}}],
\end{align}
and the orbital magnetic moment
\begin{align}
\bm{m}_{\bm k}= -\frac{e}{2\hbar} \t{Im}[\bra{\nabla_{\bm k} u_{\bm k}} \times (H_{\bm k} -\epsilon^0_{\bm k}) \ket{\nabla_{\bm k} u_{\bm k}}].
\end{align}
with the convention for the electron charge $e>0$, and $\bra{u_{\bm k}}$ is Bloch's wavefunction of a single band.
When $\bm B \neq 0$, the magnetic orbital moment shifts the energy dispersion in the absence of magnetic field, $\epsilon^0_{\bm k}$, to  
\begin{equation}
\label{eq:gendisp}
\epsilon_{\bm k}=\epsilon^0_{\bm k}-\bm{m}_{\bm k}\cdot \bm{B}
\end{equation}
where $H_{\bm k} \ket{u_{\bm k}}=\epsilon^0_{\bm k} \ket{u_{\bm k}}$

The current density $\bm j$ is given by
\begin{equation}
\label{eq:current}
\bm j =-e\int [d\bm k] (D \dot{\bm r}  + \bm{\nabla_r} \times \bm{m}_{\bm k} )f,
\end{equation}
with $[d\bm k] =d\bm k/(2\pi)^3$ and $f$ the field-induced distribution function.
The second term on the right-hand side only contributes to generate magnetization currents.
As our experiment only measures transport currents, we drop it from now on.
The factor
\begin{equation}
    D=1+ \frac{e}{\hbar} \bm B \cdot \bm \Omega_{\bm k},
\end{equation}
takes into account the change of the volume of the phase space in an applied magnetic field \cite{Xiao2010}.
It is convenient to rewrite the equations of motion \eqref{eq: EOM} as
\begin{subequations}
\begin{align}
\label{eq:rdot}
\dot{\bm r}&=\frac{1}{\hbar D}[
\bm{\nabla_k} \epsilon_{\bm k} + e{\bm E} \times \bm \Omega_{\bm k} + \frac{e}{\hbar}(\bm{\nabla_k} \epsilon_{\bm k} \cdot \bm \Omega_{\bm k})\bm B], \\
\label{eq:kdot}
\hbar \dot{\bm k}&=\frac{1}{D}[
-e{\bm E} - \frac{e}{\hbar} \bm{\nabla_k} \epsilon_{\bm k} \times \bm B - \frac{e ^2}{\hbar}(\bm{E} \cdot \bm B)\bm \Omega_{\bm k}]
\end{align}
\end{subequations}
Using first equation above we can write \eqref{eq:current} as
\begin{align}
\bm j &=-e \int [d\bm k] [
\bm{\tilde v}_{\bm k} + \frac{e}{\hbar} \bm E \times \bm \Omega_{\bm k} + \frac{e}{\hbar}(\bm{\tilde v}_{\bm k} \cdot \bm \Omega_{\bm k})\bm B] f,
\label{eq: j with B}
\end{align}
where we defined the generalized velocity
\begin{align}
\bm{\tilde v}_{\bm k}= \bm{v}_{\bm k}-(1/\hbar)\bm{\nabla_{k}}(\bm m \cdot \bm B),
\end{align}
in terms of the zero-field velocity $\bm{v}_{\bm k}=(1/\hbar)\bm{\nabla_{k}}\epsilon_{\bm k}^0$.

The Boltzmann equation determines the Fermi distribution function in terms of the fields.
In the steady state, and for a uniform system we may write it as
\begin{align}
\bm{\dot{k}} \cdot \bm{\nabla_k} f= I_\mathrm{coll}.
\label{eq:Boltzmanneq}
\end{align}
For the right-hand side we assume the relaxation time approximation, where the collision integral is \cite{Mandal2022}
\begin{equation}
   I_\mathrm{coll}= -\frac{f-f_0}{\tau} +\left(1-\dfrac{\tau}{\tau_v}\right)\frac{\delta\mu}{\tau}\partial_\epsilon f_0.
\end{equation}
Here, $\tau_v$ is the internode scattering and $\tau$ is the intranode scattering, while $\delta \mu$ accounts for the potential chemical potential shift between the two valleys, as in Ref.~\cite{Mandal2022}.
We will assume that $\delta\mu=c_a\mathbf{E}\cdot \mathbf{B}$, where $c_a$ is a chiral anomaly coefficient proportional to the internode scattering time $\tau_v$~\cite{Mandal2022}.

We are interested in finding a solution of the form $f=f_0+f_1$ such that 
\begin{equation}
\label{eq:FD}
 f_0(\epsilon_{\bm k}-E_F)=\theta(E_F-\epsilon^0_{\bm k}-\bm{m_k}\cdot \bm B)   
\end{equation}
is the equilibrium distribution function for a Fermi energy $E_F$ at $T=0$, defined by the step-function $\theta(x)=0 (x<0), 1 (x \ge 0)$. We emphasize that $\epsilon_{\bm k}$ depends on magnetic field already through \eqref{eq:gendisp}.

Solving \eqref{eq:Boltzmanneq} using \eqref{eq:kdot} we obtain
\begin{eqnarray}\nonumber
&-\dfrac{e}{D}&[{\bm E} +\frac{e}{\hbar}(\bm{E} \cdot \bm B)\bm \Omega_{\bm k}] \cdot \bm{\nabla_p} f_0 -\dfrac{e}{D} \left(\bm{v_k} \times \bm B\right)\cdot\bm{\nabla_p} f_1\\
&=&
\label{boltz2}
-\frac{f_1}{\tau} +\left(1-\dfrac{\tau}{\tau_v}\right)\frac{\delta\mu}{\tau}\partial_\epsilon f_0,
\end{eqnarray}
where we defined for convenience $\bm{\nabla_p}=(1/\hbar)\bm{\nabla_k}$.
We note that $\bm{\nabla_p} f_0= (1/\hbar) (\bm{\nabla_k} \epsilon_{\bm k}) \partial_\epsilon f_0$.
We can solve Eq.~\eqref{boltz2} recursively as \cite{Mandal2022}
\begin{eqnarray}
\nonumber
f_1 &=&\sum_{i=0} \left(\dfrac{\tau e}{D}\right)^{i}\hat{L}^{i}\left[\left(\dfrac{\tau e}{D}\right)\left({\bm E} +\frac{e}{\hbar}(\bm{E} \cdot \bm B)\bm \Omega_{\bm k}\right)\right] \cdot \bm{\nabla_p} f_0\\
\label{eq:solbol}
&-&\left(1-\dfrac{\tau}{\tau_v}\right){\delta\mu}\partial_\epsilon f_0,
\end{eqnarray}
where $\hat{L}=\left(\bm{v_k} \times \bm B\right)\cdot \bm{\nabla_p}$ is the Lorentz operator. 
We are now in position to write the current to linear order in the electric field.
We separate the current terms in powers of the scattering time up to second order
\begin{eqnarray}
\label{eq:js}
\bm{j}&=& \bm{j}_0+\bm{j}_1+\bm{j}_2\\
\label{eq:tau0}
\bm{j}_0&=&
-e  \int_\t{BZ} [d\bm k]
\frac{e}{\hbar} {\bm E} \times \bm \Omega_{\bm k} f_0\\
\label{eq:tau1}
\bm{j_1}&=&-e \tau \int_\t{BZ} [d\bm k]\Big\{\frac{1}{D}
\left[
\bm{\tilde v}_{\bm k} 
+\frac{e}{\hbar}(\bm \Omega_{\bm k}\cdot \bm{\tilde v}_{\bm k} )\bm B
\right]
\left[
 (e\bm{\tilde v}_{\bm k} 
 +\dfrac{\tau_c}{\tau}\frac{e ^2}{\hbar}(\bm \Omega_{\bm k}\cdot \bm{\tilde v}_{\bm k}) \bm B) \cdot {\bm E}  \partial_\epsilon f_0
 \right]
\Big\},\\
\label{eq:tau2}
\bm{j_2}&=&-e \tau^2 \int_\t{BZ} [d\bm k]\Big\{\frac{1}{D}
\left[
\bm{\tilde v}_{\bm k} 
+\frac{e}{\hbar}(\bm \Omega_{\bm k}\cdot \bm{\tilde v}_{\bm k} )\bm B
\right]
\left[
\frac{1}{D}\hat{L}
 (e\bm{\tilde v}_{\bm k} 
 +\frac{e ^2}{\hbar}(\bm \Omega_{\bm k}\cdot \bm{\tilde v}_{\bm k}) \bm B) \cdot {\bm E}  \partial_\epsilon f_0
 \right]
\Big\},
\end{eqnarray}

The term $\bm{j}_0$ encodes a Hall effect.
Because Eq.~\eqref{eq:FD} depends on magnetic field and the orbital moment, the Hall effect in  $\bm{j}_0$ has different contributions.
At zero magnetic field, the dependence on the orbital magnetization drops and $\bm{j}_1$ is only non-zero if the material breaks-time reversal symmetry.
Such anomalous Hall effect vanishes for time-reversal symmetric systems such as CoSi.
At finite magnetic field, $\bm{j}_0$ vanishes only in the absence of orbital magnetization, even when the material respects time-reversal symmetry.
When inversion symmetry is broken, a nonzero orbital magnetization generates an intrinsic Hall current proportional to $\bm B$, derived in Ref.~\cite{Gao2014} and later calculated for a single Weyl node in Ref.~\cite{Agarwal21}.
This contribution, which we call orbital-magnetic-moment Hall effect, arises from a $k$-asymmetry in the energy spectrum due to the magnetic field, as given by \eqref{eq:gendisp}~\cite{Cai2013}.

The term $\bm{j}_1$ contains terms proportional to a scattering time. 
These will result in a longitudinal magnetoresistance, and will include Drude contributions, Berry curvature contributions, and orbital magnetic moment contributions.
To obtain the last term in $\bm{j}_1$, we noticed that $\delta\mu$ can be found self-consistently to be proportional to $\mathbf{E}\cdot \mathbf{B}$, since $\delta\mu= c \tau_v \mathbf{E}\cdot \mathbf{B}$, where $c$ is a constant~\cite{Deng2019,Mandal2022}.
Hence, the terms contributing to $\bm{j}_1$ coming from the second line in Eq.~\eqref{eq:solbol} and the last term in the first line in Eq.~\eqref{eq:solbol} can be combined into a single term by redefining a new scattering time $\tau_c$. 
We call $\tau_c$ the chiral scattering time, which in general depends on both the inter and intranode scattering.
When $\tau \ll \tau_v$, $\tau_c \approx \tau_v$.
This assumption is well justified since the multi-fold fermions are maximally separated in momentum space, and we will adopt it in what follows.
We determine the ratio $\tau_v/\tau$ experimentally, as discussed in the main text and below.

The term $\bm{j}_2$, proportional to $\tau^2$, includes the classical Hall effect due the Lorentz force.
We now calculate all contributions for the multifold fermions of CoSi.

\subsection{Angle-dependent magnetoconductance of CoSi}

At the $R$ point, and neglecting spin-orbit coupling, as reasonable for CoSi~\cite{chang_unconventional_2017,Ni:2021cta}, there are two Weyl points of the same chirality, doped above the crossing point ($\mu>0$).
The Hamiltonian to linear order in momentum for such a \textit{double Weyl} fermion reads
\begin{align}
H^{(dW)}= \chi_R v_R \bm \sigma \cdot \bm k \otimes \tau_0 \otimes s_0,
\label{eq: H Weyl}
\end{align}
where $v_R$ is the Fermi velocity at the $R$ point, $\chi_R=\pm1$ is the chirality of the $R$ node and $\sigma$ is a vector of Pauli matrices representing an orbital-degree of freedom. 
Using $\tau_0$ we encode the valley degeneracy, i.e. the $R$ point is composed of two Weyls of the same chirality, and $s_0$ we encode the spin degeneracy.
For a single Weyl fermion, the Berry curvature and orbital magnetic moment of each of the bands are 
\begin{subequations}
\label{eq: Omega Weyl}
\begin{align}
\bm \Omega_{\bm k} &= -\chi_R\frac{s}{2 k^2} \bm{\hat k}, \\
\bm m_{\bm k} &=-\chi_R\frac{ev_R}{2 k} \bm{\hat k},
\end{align}
\end{subequations}
where $\hat{\bm{k}}$ is the unit vector along $\bm k$, and $s=-1,+1$ for the valence and conduction band, respectively. While the Berry curvature changes sign for the each band, the orbital magnetic moment does not.

Consistent with all experimental reports of CoSi samples~\cite{takane_observation_2019,Xu:2020ei,Ni:2021cta} and our quantum oscillations data we assume that R point is filled above the node up to its conduction band, with Fermi momentum $k^R_F$, and we assign it a chirality $\chi_R = +1$.

We assume an electric field along the $x$ direction, 
$\bm E =(E_x,0,0)$, and vary the magnetic field by an angle $\theta$ with respect to the $x$ direction, $\bm B =B_0 (\cos\theta,0,\sin\theta)$.
The longitudinal conductivity for the component of the current along the electric field for the double Weyl at $R$ is entirely due to Eq.~\eqref{eq:tau1}.
The corresponding current density $j^{(dW)}_x$ and conductivity $ \sigma^{(dW)}_{xx}$ read
\begin{subequations}
\begin{align}
    j^{(dW)}_x &= \sigma^{(dW)}_{xx} E_x \\
\label{eq:jdW2}
    \sigma^{(dW)}_{xx} &= 4\left[\frac{\tau e^2 v_{R}(k^{R}_F)^2}{6 \pi^2 \hbar}   
+ \frac{\left[(1+\frac{5\tau_v}{\tau})\cos^2\theta-2\right]\tau e^4 v_{R} B_0^2}{120 \pi^2 \hbar^3 (k^{R}_F)^2} \right]
\end{align}    
\end{subequations}
The factor $4$ in front of the squared brackets takes into account the existence of two Weyl nodes at $R$ and the spin degeneracy.
As we sweep the angle from $\theta=0$ to $\theta=\pi/2$ the longitudinal magnetoconductance
goes from being positive to negative due to the second term on the right-hand side.
The chiral anomaly contributes to this term via the last term in Eq.~\eqref{eq:tau1}, via the term proportional to $\tau_v/\tau$. It has a $\cos^2\theta$ angular dependence as observed in our experiment.
Later we will fix $\tau_v/\tau$ with our experiments, which suggest  $\tau_v/\tau \gg 1$.
However, if we assume $\tau=\tau_v$ Eq.\eqref{eq:jdW2} recovers the results of Ref.~\cite{Morimoto2016}, consistent with the fact that this reference did not distinguish inter and intranode scattering. 
\\

To linear order in momentum, the \textit{threefold fermion} at $\Gamma$ is described by the Hamiltonian
\begin{align}
H^{(3f)}= \chi_\Gamma v_\Gamma \bm S \cdot \bm k \otimes s_0,
\label{eq: threefold}
\end{align}
where $v_\Gamma$ is the Fermi velocity, $\bm S$ is a vector of spin-1 matrices representing the orbital-degree of freedom, and  $\chi_\Gamma=\pm1$ is the chirality of the $\Gamma$ node. Using $s_0$ we encode again the spin degeneracy.
Eq.~\eqref{eq: threefold} needs to describe a threefold Fermion with the opposite chirality compared to \eqref{eq: H Weyl}, as enforced by the Nielsen-Ninomiya, or fermion doubling, theorem~\cite{NielNino81a,NielNino81b}, \textit{i.e.} $\chi_\Gamma=-\chi_R$. 

The threefold Hamiltonian has three bands which are spin-degenerate.
The corresponding Berry curvature and orbital magnetic were calculated in Ref.~\cite{Flicker2018} and are 
\begin{subequations}
\begin{align}
\bm \Omega^{n}_{\bm k} &= \eta_\Omega^{n}\chi_\Gamma \frac{1}{2 k^2} \bm{\hat k}, \\
\bm m^{n}_{\bm k} &= \eta_m^{n}\chi_\Gamma \frac{ev_\Gamma}{2 k} \bm{\hat k},
\end{align}
\label{eq:Omega3f}
\end{subequations}
where $n$ labels the bottom, middle and upper band of the threefold fermion, $n=1,2,3$ respectively.
Unlike the double-Weyl node, where the Chern numbers of the band are $2\times(1,-1)$, the threefold bands have Chern numbers $2,0,-2$. 
Hence $\eta_\Omega^{n}=(-2,0,2)$ for the top, middle and bottom bands, respectively.
The orbital magnetic moment of top and bottom bands equals that of a Weyl fermion. For the middle band, the orbital magnetic moment is twice that of a Weyl fermion. This results in $\eta_m^{n}=(1,2,1)$~\cite{Flicker2018}.

\begin{figure*}[htbp]
\includegraphics[width=\textwidth]{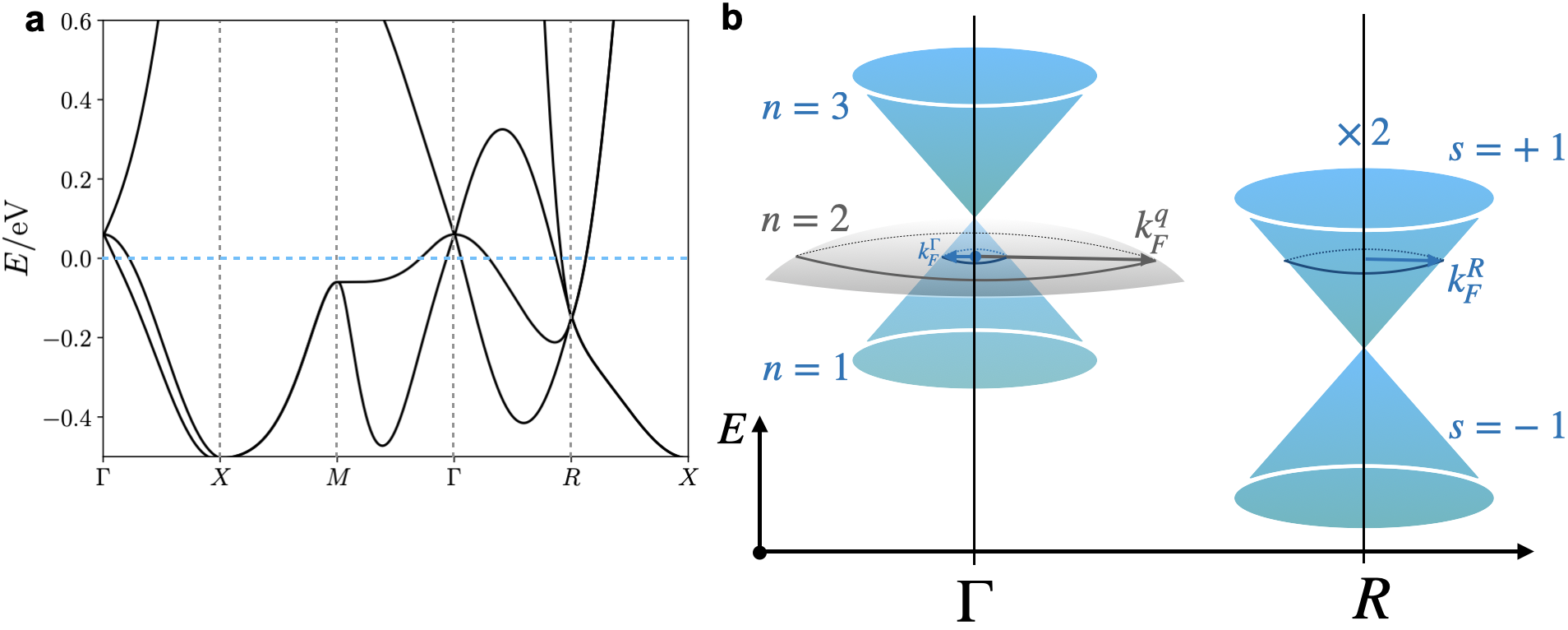}
\caption{\label{fig:SI_MF}{\textbf{Low energy semiclassical theory.} {\bf{a}} Tight-binding band structure energy versus momentum of CoSi crystal. {\bf{b}} Schematic of energy versus momentum of the threefold fermion located at $\Gamma$ and the fourfold fermion located at $R$ (two Weyl cones of equal chirality). In our semiclassical calculation we assume that $k^q_F,k^R_F\gg k^\Gamma_F$, consistent with previous experiments~\cite{takane_observation_2019,Xu:2020ei,Ni:2021cta} and our quantum-oscillation data. Ignoring spin, $n=1,2,3$ labels the three bands at $\Gamma$, while $s=-1,1$ labels the valence and conduction bands at $R$, respectively, each doubly degenerate (indicated with a $\times 2$ in the figure). Their Berry curvatures and orbital magnetic moments are given by Eqs.~\eqref{eq: Omega Weyl} and \eqref{eq:Omega3f}. Considering spin we add an additional two-fold degeneracy for all bands at $\Gamma$ and $R$. }}
\end{figure*}

Next, we assume that the chemical potential crosses the lower two bands,
the linearly dispersing band ($n=1$) and the central quadratic band ($n=2$), \textit{i.e.} we assume $\mu_{3f}<0$ measured with respect to the threefold crossing. We label the corresponding Fermi momentum of the linear band $k^{\Gamma}_F$. Since the middle band is flat in the linear approximation, we promote it to a quadratic, hole-like band, for it to cross the Fermi level. 
We call its Fermi momentum $k^{q}_F\gg k^{\Gamma}_F$, see Fig.~\ref{fig:SI_MF}.
These assumptions are justified by previously reported density functional theory (DFT) calculations  and ARPES experiments~\cite{takane_observation_2019,Xu:2020ei,Ni:2021cta}.

We first calculate the magneto-conductivity of the $n=1$ setting the chirality in \eqref{eq: threefold} to be the opposite of that of R, i.e. $\chi_\Gamma=-1$. Assuming the same configuration of electric and magnetic fields as above and using Eq.~\eqref{eq:tau1} the component of the current along the electric field for the threefold fermion at $\Gamma$ is
\begin{subequations}
\begin{align}
    j^{(3f)}_x &= \sigma^{(3f)}_{xx} E_x,\\
 \sigma^{(3f)}_{xx} &= 
 2\left[\frac{\tau e^2 v_{\Gamma}(k^{\Gamma}_F)^2}{6 \pi^2 \hbar}   
+ \frac{\left[(-9+30\frac{\tau_v}{\tau})\cos^2\theta-2\right]\tau e^4 v_{\Gamma} B_0^2}{120 \pi^2 \hbar^3 (k^{\Gamma}_F)^2} \right]. 
 \end{align}
\end{subequations}
The factor 2 in front of this expression takes into account the spin degeneracy.
As a benchmark, we note that setting the magnetic moment to zero and $\theta=0$ we recover the expression given by \cite{Nandy2019}. 
However, as we emphasized in the main text, since the magnetic moment is as sizable as the Berry curvature, it cannot be neglected.

Lastly, using the same formalism we can calculate the contribution of the  central parabolic band.
As it is a quadratic band with no Chern number, it is expected to lead to an ordinary negative magnetoconductance (positive magnetoresistance).
Its magnitude is inversely proportional to the Fermi momentum $k^q_F$ of the quadratic band, i.e. $\sigma \propto \frac{e^4 \tau  (v_\Gamma)^2 B^2_0}{\hbar^2E_Fk^q_F} $, where  $v_\Gamma$ is the Fermi velocity of the linear bands.
Compared to the linearly dispersing bands, the quadratic band has a relatively large $E_F$ and $k^q_F$, which allows us to neglect its positive contribution to magneto-conductivity.

Defining the magneto-conductance as
\begin{equation}
    \sigma_{PMC}(B_0)\equiv\frac{\sigma(B_0) -\sigma(B_0 = 0)}{\sigma(B_0=0)}
\end{equation}
Combining the contribution from both multifold fermions we have
\begin{widetext}
\begin{equation}
\sigma_{PMC}(B_0)=
\frac{\left[
\left(-9+2 r_1
+10\left(3+r_1\right)
\dfrac{\tau_v}{\tau}
\right) \cos^2\theta- \left(2+4r_1\right)\right]}{
20\left(1+2r_2 \right) }\, \dfrac{B_0^2}{\left(k^\Gamma_F\right)^4}\dfrac{e^2}{\hbar^2}.
\label{eq:all}
\end{equation}
\end{widetext}
with $r_1=\left(\frac{k^{\Gamma}_F}{k^{R}_F}\right)^2\frac{v_{R}}{v_{\Gamma}}$
and $r_2=\left(\frac{k^{R}_F}{k^{\Gamma}_F}\right)^2\frac{v_{R}}{v_{\Gamma}}$, resulting in Eq.(1) in the main text.
To estimate the positive magnetoconductance we can use that tight-binding models fit well the DFT dispersion if $v_R=v_\Gamma/ \sqrt{3}$ and that from quantum oscillations $k^R_F \approx 5 k^\Gamma_F$.
In the longitudinal configuration ($\theta=0$) this simplifies to
%
{
\begin{equation}
\sigma_{PMC}(B_0)\Big\vert_{\theta=0} \approx \frac{\left[-825-2\sqrt{3}+10\left(225 +\sqrt{3}\right)\dfrac{\tau_v}{\tau}\right] e^2 B_0^2}{
500 \left(3+50\sqrt{3}\right)\hbar^2\left(k^\Gamma_F\right)^4 }\, .
\label{eq:allsimp}
\end{equation}
}

We can use 
$k^\Gamma_F=0.25 \mathrm{nm}^{-1}$ to get

\begin{equation}
\sigma_{PMC}(B_0=1T,\theta=0) \approx -0.001 \% +0.003\%\dfrac{\tau_v}{\tau}.
\label{eq:allsimp2}
\end{equation}

We note that when the internode scattering time is slower than the intranode scattering time, $\frac{\tau_v}{\tau}\ll 1$, the magnetoconductance becomes negative. 
This is consistent with the physical intuition that the anomalous positive magnetoresistance occurs only when internode scattering dominates, i.e. $\frac{\tau_v}{\tau}\gg 1$.
By treating $\frac{\tau_v}{\tau}$ as a fitting parameter, we can include other trivial scattering terms that we have neglected in the calculation, e.g. between Weyl and trivial bands that lead to positive magnetoresistance~\cite{suh2024effect}.

With the above we need that $\frac{\tau_v}{\tau}\approx 10^2$ to get an enhancement of 1$\%$ at 1T as seen experiment. That $\frac{\tau_v}{\tau}\gg 1$ is reasonable, since the chiral multifold fermions are maximally separated in the Brillouin zone. 

We have checked that for both the $\Gamma$ and the $R$ points the orbital magnetic moment terms always contribute to decrease the magneto-conductance, \textit{i.e}. they result in positive magnetoresistance. Therefore, their effect is to reduce the chiral anomaly negative magnetoresistance, but their magnitude is insufficient to overcome it.

\section{non-linear contribution to the Hall effect}

This section contains the main theoretical results behind our interpretation of the Hall conductivity.
The large orbital magnetic moment and Berry curvature close to the node at $\Gamma$ are the main contributors to a cubic-in-magnetic field contribution to the Hall effect.  
This result is a consequence of the large orbital magnetic moment and Berry curvature close to the $\Gamma$ point.

We wish to argue why we can expect the cubic terms of the Hall effect to follow the positive magneto-conductivity, i.e. magnetic fields where we observe positive magnetoconductance have also an observable cubic contribution to the Hall effect. 

First let us discuss why we expect non-linear Hall terms in the semiclassical picture.
The Hall response is determined solely by the intrinsic (i.e. $\tau$ independent) Eq.~\eqref{eq:tau0} and the Lorentz part Eq.~\eqref{eq:tau2}, proportional to $\tau^2$.
The dominant linear-in-$B$ term, that we subtract in the main text to highlight the cubic contribution, is due to the Lorentz force.
It is proportional to the density of carriers, and it entirely determined by Eq.~\eqref{eq:tau2}.
After the linear-in-$B$ contribution is subtracted, both Eq.~\eqref{eq:tau0} and Eq.~\eqref{eq:tau2} contribute to the cubic $B^3$-dependent part of the Hall conductivity.

In the following we will again assume $\bm E =(E_x,0,0)$, and vary the magnetic field by an angle $\theta$ with respect to the $x$ direction, $\bm B =B_0 (\cos\theta,0,\sin\theta)$. Contrary to the previous section, we will here be interested in the $y$ component of the current, $j_y$, which is perpendicular to $\bm E $. We now analyze the separate contributions and determine the dominant cubic terms.

\subsection{Threefold at $\Gamma$}

The three-fold fermion at the $\Gamma$ point has three bands, a central quadratic band that is relatively flat (has a large effective mass), and two linearly dispersing bands. The central quadratic band has zero Berry curvature and finite orbital angular momentum while the linearly dispersing bands have both Berry curvature and finite orbital angular momentum, as per Eq.~\eqref{eq:Omega3f}. Consistent with our quantum oscillation data, previous photoemission~\cite{takane_observation_2019} and optical experiments~\cite{Xu:2020ei,Ni:2021cta} we assume that the lowest ($n=1$) and central ($n=2$) $\Gamma$ bands are partially occupied, with momentum $k^{\Gamma}_F$ and $k^q_F$ respectively (see Fig.~\ref{fig:SI_MF}).

The bottom linear band contributes through Eq.~\eqref{eq:tau2}
\begin{equation}
\label{eq:cubic1}
    j_{2,y}=-\dfrac{(e^3 \sin\theta( e^2 B_0^2 + 20 \hbar^2 (k^\Gamma_F)^4) \tau^2 v_\Gamma^2)B_0E_x}{60\pi^2 \hbar^4 (k^\Gamma_F)^3},
\end{equation}
and through Eq.~\eqref{eq:tau0}
\begin{equation}
\label{eq:cubic2}
    j_{0,y}=- \dfrac{e^3 (9 B_0^2 e^2 + 
   5 \hbar^2 (k^\Gamma_F)^4)  B_0E_x\sin\theta }{30\pi^2 \hbar^4 (k^\Gamma_F)^5 },
\end{equation}
where we have taken into account the spin degeneracy.

The central quadratic band at $\Gamma$ has zero orbital magnetic moment and which, combined with time-reversal symmetry, renders $j_{0,y}=0$, from Eq.~\eqref{eq:tau0}.
Through Eq.~\eqref{eq:tau2} this band contributes 
%
\begin{equation}
    j_{2,y}=-\dfrac{ e^3 v^2_\Gamma \tau^2 \left(\frac{4(E^q_F k^q_F)^2 }{v_\Gamma^2} + 
   B_0^2 e^2 \right) B_0E_x\sin\theta}{3 \pi^2 \hbar^4 (k^q_F)^3}.
\end{equation}
where, as we did previously, we assumed a hole-like parabolic dispersion with Fermi momentum $k^q_F$ and Fermi energy $E^q_F$, and have taken into account the spin degeneracy.

\subsection{Fourfold at $R$}
The two top linear bands contribute through Eq.~\eqref{eq:tau2} as
\begin{equation}
    j_{2,y}=\dfrac{-4\tau^2e^3v^2_R \left(    
    e^2 B^2_0-5\hbar^2(k^R_F)^4 \right)B_0E_x\sin\theta}{30\pi^2\hbar^4(k^R_F)^3},
\end{equation}
and the contribution from Eq.~\eqref{eq:tau0} is
\begin{equation}
    j_{0,y}=\dfrac{-4e^3 \left(9e^2B^2_0+5\hbar^2(k^R_F)^4\right)B_0E_x\sin\theta}{120\pi^2\hbar^4(k^R_F)^5}.
\end{equation}
As with the longitudinal conductivity, the factor 4 takes into account the double Weyl node structure and the spin degeneracy.
\subsection{Dominant contribution to the cubic terms in the Hall conductivity}

In the main text we subtract the linear Hall conductivity and are left with a cubic contribution $\sigma_{xy}\propto B_0^3$.
Experimentally, the central quadratic band at $\Gamma$ and the R-point double Weyl form the largest Fermi-surfaces (see discussion in main text, Fig. 1a, and e.g.\cite{takane_observation_2019,Huber2023}). 
Hence, we have $k^q_F,k^R_F\gg k^\Gamma_F$. 
With this condition, assuming the same order of magnitude for $\tau$ in all bands, the main contribution to the cubic term in $\sigma_{xy}$ comes from Eqs.~\eqref{eq:cubic1} and ~\eqref{eq:cubic2}. Both of these equations are finite due to the orbital magnetic moment and the Berry curvature of the linear band around $\Gamma$.
Since these quantities diverge close to the multifold crossing point at $\Gamma$, see Eq.~\eqref{eq:Omega3f}, we conclude that the dominant contribution to the cubic term in the non-linear Hall is the filled linear band at $\Gamma$. 
The positive magnetoconductance is also determined by the Berry curvature and orbital magnetic moment close to the node. 
Hence, it is reasonable that the cubic Hall terms and positive magneto-conductance are sizable around the same values of magnetic field, as observed in our experimental data.


\begin{thebibliography}{63}%
\makeatletter
\providecommand \@ifxundefined [1]{%
 \@ifx{#1\undefined}
}%
\providecommand \@ifnum [1]{%
 \ifnum #1\expandafter \@firstoftwo
 \else \expandafter \@secondoftwo
 \fi
}%
\providecommand \@ifx [1]{%
 \ifx #1\expandafter \@firstoftwo
 \else \expandafter \@secondoftwo
 \fi
}%
\providecommand \natexlab [1]{#1}%
\providecommand \enquote  [1]{``#1''}%
\providecommand \bibnamefont  [1]{#1}%
\providecommand \bibfnamefont [1]{#1}%
\providecommand \citenamefont [1]{#1}%
\providecommand \href@noop [0]{\@secondoftwo}%
\providecommand \href [0]{\begingroup \@sanitize@url \@href}%
\providecommand \@href[1]{\@@startlink{#1}\@@href}%
\providecommand \@@href[1]{\endgroup#1\@@endlink}%
\providecommand \@sanitize@url [0]{\catcode `\\12\catcode `\$12\catcode `\&12\catcode `\#12\catcode `\^12\catcode `\_12\catcode `\%12\relax}%
\providecommand \@@startlink[1]{}%
\providecommand \@@endlink[0]{}%
\providecommand \url  [0]{\begingroup\@sanitize@url \@url }%
\providecommand \@url [1]{\endgroup\@href {#1}{\urlprefix }}%
\providecommand \urlprefix  [0]{URL }%
\providecommand \Eprint [0]{\href }%
\providecommand \doibase [0]{https://doi.org/}%
\providecommand \selectlanguage [0]{\@gobble}%
\providecommand \bibinfo  [0]{\@secondoftwo}%
\providecommand \bibfield  [0]{\@secondoftwo}%
\providecommand \translation [1]{[#1]}%
\providecommand \BibitemOpen [0]{}%
\providecommand \bibitemStop [0]{}%
\providecommand \bibitemNoStop [0]{.\EOS\space}%
\providecommand \EOS [0]{\spacefactor3000\relax}%
\providecommand \BibitemShut  [1]{\csname bibitem#1\endcsname}%
\let\auto@bib@innerbib\@empty
\bibitem [{\citenamefont {Armitage}\ \emph {et~al.}(2018)\citenamefont {Armitage}, \citenamefont {Mele},\ and\ \citenamefont {Vishwanath}}]{Armitage18}%
  \BibitemOpen
  \bibfield  {author} {\bibinfo {author} {\bibfnamefont {N.~P.}\ \bibnamefont {Armitage}}, \bibinfo {author} {\bibfnamefont {E.~J.}\ \bibnamefont {Mele}},\ and\ \bibinfo {author} {\bibfnamefont {A.}~\bibnamefont {Vishwanath}},\ }\bibfield  {title} {{\selectlanguage {English}\bibinfo {title} {{Weyl and Dirac semimetals in three-dimensional solids}}},\ }\href {https://doi.org/10.1103/revmodphys.90.015001} {\bibfield  {journal} {\bibinfo  {journal} {Reviews of Modern Physics}\ }\textbf {\bibinfo {volume} {90}},\ \bibinfo {pages} {015001} (\bibinfo {year} {2018})}\BibitemShut {NoStop}%
\bibitem [{\citenamefont {Manes}(2012)}]{manes_existence_2012}%
  \BibitemOpen
  \bibfield  {author} {\bibinfo {author} {\bibfnamefont {J.~L.}\ \bibnamefont {Manes}},\ }\bibfield  {title} {\bibinfo {title} {Existence of bulk chiral fermions and crystal symmetry},\ }\bibfield  {journal} {\bibinfo  {journal} {Physical Review B}\ }\textbf {\bibinfo {volume} {85}},\ \href {https://doi.org/10.1103/PhysRevB.85.155118} {10.1103/PhysRevB.85.155118} (\bibinfo {year} {2012})\BibitemShut {NoStop}%
\bibitem [{\citenamefont {Bradlyn}\ \emph {et~al.}(2016)\citenamefont {Bradlyn}, \citenamefont {Cano}, \citenamefont {Wang}, \citenamefont {Vergniory}, \citenamefont {Felser}, \citenamefont {Cava},\ and\ \citenamefont {Bernevig}}]{Bradlyn2016}%
  \BibitemOpen
  \bibfield  {author} {\bibinfo {author} {\bibfnamefont {B.}~\bibnamefont {Bradlyn}}, \bibinfo {author} {\bibfnamefont {J.}~\bibnamefont {Cano}}, \bibinfo {author} {\bibfnamefont {Z.}~\bibnamefont {Wang}}, \bibinfo {author} {\bibfnamefont {M.~G.}\ \bibnamefont {Vergniory}}, \bibinfo {author} {\bibfnamefont {C.}~\bibnamefont {Felser}}, \bibinfo {author} {\bibfnamefont {R.~J.}\ \bibnamefont {Cava}},\ and\ \bibinfo {author} {\bibfnamefont {B.~A.}\ \bibnamefont {Bernevig}},\ }\bibfield  {title} {\bibinfo {title} {Beyond dirac and weyl fermions: Unconventional quasiparticles in conventional crystals},\ }\href {https://doi.org/10.1126/science.aaf5037} {\bibfield  {journal} {\bibinfo  {journal} {Science}\ }\textbf {\bibinfo {volume} {353}},\ \bibinfo {pages} {aaf5037} (\bibinfo {year} {2016})},\ \Eprint {https://arxiv.org/abs/https://www.science.org/doi/pdf/10.1126/science.aaf5037} {https://www.science.org/doi/pdf/10.1126/science.aaf5037} \BibitemShut {NoStop}%
\bibitem [{\citenamefont {Tang}\ \emph {et~al.}(2017)\citenamefont {Tang}, \citenamefont {Zhou},\ and\ \citenamefont {Zhang}}]{tang_multiple_2017}%
  \BibitemOpen
  \bibfield  {author} {\bibinfo {author} {\bibfnamefont {P.}~\bibnamefont {Tang}}, \bibinfo {author} {\bibfnamefont {Q.}~\bibnamefont {Zhou}},\ and\ \bibinfo {author} {\bibfnamefont {S.-C.}\ \bibnamefont {Zhang}},\ }\bibfield  {title} {\bibinfo {title} {Multiple types of topological fermions in transition metal silicides},\ }\bibfield  {journal} {\bibinfo  {journal} {Physical Review Letters}\ }\textbf {\bibinfo {volume} {119}},\ \href {https://doi.org/10.1103/PhysRevLett.119.206402} {10.1103/PhysRevLett.119.206402} (\bibinfo {year} {2017}),\ \bibinfo {note} {arXiv: 1706.03817}\BibitemShut {NoStop}%
\bibitem [{\citenamefont {Chang}\ \emph {et~al.}(2018)\citenamefont {Chang}, \citenamefont {Wieder}, \citenamefont {Schindler}, \citenamefont {Sanchez}, \citenamefont {Belopolski}, \citenamefont {Huang}, \citenamefont {Singh}, \citenamefont {Wu}, \citenamefont {Chang}, \citenamefont {Neupert}, \citenamefont {Xu}, \citenamefont {Lin},\ and\ \citenamefont {Hasan}}]{Chang:2018bb}%
  \BibitemOpen
  \bibfield  {author} {\bibinfo {author} {\bibfnamefont {G.}~\bibnamefont {Chang}}, \bibinfo {author} {\bibfnamefont {B.~J.}\ \bibnamefont {Wieder}}, \bibinfo {author} {\bibfnamefont {F.}~\bibnamefont {Schindler}}, \bibinfo {author} {\bibfnamefont {D.~S.}\ \bibnamefont {Sanchez}}, \bibinfo {author} {\bibfnamefont {I.}~\bibnamefont {Belopolski}}, \bibinfo {author} {\bibfnamefont {S.-M.}\ \bibnamefont {Huang}}, \bibinfo {author} {\bibfnamefont {B.}~\bibnamefont {Singh}}, \bibinfo {author} {\bibfnamefont {D.}~\bibnamefont {Wu}}, \bibinfo {author} {\bibfnamefont {T.-R.}\ \bibnamefont {Chang}}, \bibinfo {author} {\bibfnamefont {T.}~\bibnamefont {Neupert}}, \bibinfo {author} {\bibfnamefont {S.-Y.}\ \bibnamefont {Xu}}, \bibinfo {author} {\bibfnamefont {H.}~\bibnamefont {Lin}},\ and\ \bibinfo {author} {\bibfnamefont {M.~Z.}\ \bibnamefont {Hasan}},\ }\bibfield  {title} {\bibinfo {title} {{Topological quantum properties of chiral crystals}},\ }\href {https://doi.org/10.1038/s41563-018-0169-3} {\bibfield  {journal}
  {\bibinfo  {journal} {Nature materials}\ }\textbf {\bibinfo {volume} {17}},\ \bibinfo {pages} {978} (\bibinfo {year} {2018})}\BibitemShut {NoStop}%
\bibitem [{\citenamefont {Bertlmann}(2000)}]{bertlmann2000anomalies}%
  \BibitemOpen
  \bibfield  {author} {\bibinfo {author} {\bibfnamefont {R.~A.}\ \bibnamefont {Bertlmann}},\ }\href {https://doi.org/https://doi.org/10.1093/acprof:oso/9780198507628.001.0001} {\emph {\bibinfo {title} {Anomalies in quantum field theory}}},\ Vol.~\bibinfo {volume} {91}\ (\bibinfo  {publisher} {Oxford university press},\ \bibinfo {year} {2000})\BibitemShut {NoStop}%
\bibitem [{\citenamefont {Ezawa}(2017)}]{Ezawa2017}%
  \BibitemOpen
  \bibfield  {author} {\bibinfo {author} {\bibfnamefont {M.}~\bibnamefont {Ezawa}},\ }\bibfield  {title} {\bibinfo {title} {Chiral anomaly enhancement and photoirradiation effects in multiband touching fermion systems},\ }\href {https://doi.org/10.1103/PhysRevB.95.205201} {\bibfield  {journal} {\bibinfo  {journal} {Phys. Rev. B}\ }\textbf {\bibinfo {volume} {95}},\ \bibinfo {pages} {205201} (\bibinfo {year} {2017})}\BibitemShut {NoStop}%
\bibitem [{\citenamefont {Lepori}\ \emph {et~al.}(2018)\citenamefont {Lepori}, \citenamefont {Burrello},\ and\ \citenamefont {Guadagnini}}]{Lepori2018}%
  \BibitemOpen
  \bibfield  {author} {\bibinfo {author} {\bibfnamefont {L.}~\bibnamefont {Lepori}}, \bibinfo {author} {\bibfnamefont {M.}~\bibnamefont {Burrello}},\ and\ \bibinfo {author} {\bibfnamefont {E.}~\bibnamefont {Guadagnini}},\ }\bibfield  {title} {\bibinfo {title} {Axial anomaly in multi-weyl and triple-point semimetals},\ }\href {https://doi.org/10.1007/JHEP06(2018)110} {\bibfield  {journal} {\bibinfo  {journal} {Journal of High Energy Physics}\ }\textbf {\bibinfo {volume} {2018}},\ \bibinfo {pages} {110} (\bibinfo {year} {2018})}\BibitemShut {NoStop}%
\bibitem [{\citenamefont {Nandy}\ \emph {et~al.}(2019)\citenamefont {Nandy}, \citenamefont {Manna}, \citenamefont {C\ifmmode \u{a}\else \u{a}\fi{}lug\ifmmode~\u{a}\else \u{a}\fi{}ru},\ and\ \citenamefont {Roy}}]{Nandy2019}%
  \BibitemOpen
  \bibfield  {author} {\bibinfo {author} {\bibfnamefont {S.}~\bibnamefont {Nandy}}, \bibinfo {author} {\bibfnamefont {S.}~\bibnamefont {Manna}}, \bibinfo {author} {\bibfnamefont {D.}~\bibnamefont {C\ifmmode \u{a}\else \u{a}\fi{}lug\ifmmode~\u{a}\else \u{a}\fi{}ru}},\ and\ \bibinfo {author} {\bibfnamefont {B.}~\bibnamefont {Roy}},\ }\bibfield  {title} {\bibinfo {title} {Generalized triple-component fermions: Lattice model, fermi arcs, and anomalous transport},\ }\href {https://doi.org/10.1103/PhysRevB.100.235201} {\bibfield  {journal} {\bibinfo  {journal} {Phys. Rev. B}\ }\textbf {\bibinfo {volume} {100}},\ \bibinfo {pages} {235201} (\bibinfo {year} {2019})}\BibitemShut {NoStop}%
\bibitem [{\citenamefont {Son}\ and\ \citenamefont {Spivak}(2013)}]{son_chiral_2013}%
  \BibitemOpen
  \bibfield  {author} {\bibinfo {author} {\bibfnamefont {D.~T.}\ \bibnamefont {Son}}\ and\ \bibinfo {author} {\bibfnamefont {B.~Z.}\ \bibnamefont {Spivak}},\ }\bibfield  {title} {\bibinfo {title} {Chiral anomaly and classical negative magnetoresistance of {Weyl} metals},\ }\href {https://doi.org/10.1103/PhysRevB.88.104412} {\bibfield  {journal} {\bibinfo  {journal} {Phys. Rev. B}\ }\textbf {\bibinfo {volume} {88}},\ \bibinfo {pages} {104412} (\bibinfo {year} {2013})},\ \bibinfo {note} {publisher: American Physical Society}\BibitemShut {NoStop}%
\bibitem [{\citenamefont {Arnold}\ \emph {et~al.}(2016)\citenamefont {Arnold}, \citenamefont {Shekhar}, \citenamefont {Wu}, \citenamefont {Sun}, \citenamefont {dos Reis}, \citenamefont {Kumar}, \citenamefont {Naumann}, \citenamefont {Ajeesh}, \citenamefont {Schmidt}, \citenamefont {Grushin}, \citenamefont {Bardarson}, \citenamefont {Baenitz}, \citenamefont {Sokolov}, \citenamefont {Borrmann}, \citenamefont {Nicklas}, \citenamefont {Felser}, \citenamefont {Hassinger},\ and\ \citenamefont {Yan}}]{Arnold2016}%
  \BibitemOpen
  \bibfield  {author} {\bibinfo {author} {\bibfnamefont {F.}~\bibnamefont {Arnold}}, \bibinfo {author} {\bibfnamefont {C.}~\bibnamefont {Shekhar}}, \bibinfo {author} {\bibfnamefont {S.-C.}\ \bibnamefont {Wu}}, \bibinfo {author} {\bibfnamefont {Y.}~\bibnamefont {Sun}}, \bibinfo {author} {\bibfnamefont {R.~D.}\ \bibnamefont {dos Reis}}, \bibinfo {author} {\bibfnamefont {N.}~\bibnamefont {Kumar}}, \bibinfo {author} {\bibfnamefont {M.}~\bibnamefont {Naumann}}, \bibinfo {author} {\bibfnamefont {M.~O.}\ \bibnamefont {Ajeesh}}, \bibinfo {author} {\bibfnamefont {M.}~\bibnamefont {Schmidt}}, \bibinfo {author} {\bibfnamefont {A.~G.}\ \bibnamefont {Grushin}}, \bibinfo {author} {\bibfnamefont {J.~H.}\ \bibnamefont {Bardarson}}, \bibinfo {author} {\bibfnamefont {M.}~\bibnamefont {Baenitz}}, \bibinfo {author} {\bibfnamefont {D.}~\bibnamefont {Sokolov}}, \bibinfo {author} {\bibfnamefont {H.}~\bibnamefont {Borrmann}}, \bibinfo {author} {\bibfnamefont {M.}~\bibnamefont {Nicklas}}, \bibinfo {author} {\bibfnamefont
  {C.}~\bibnamefont {Felser}}, \bibinfo {author} {\bibfnamefont {E.}~\bibnamefont {Hassinger}},\ and\ \bibinfo {author} {\bibfnamefont {B.}~\bibnamefont {Yan}},\ }\bibfield  {title} {\bibinfo {title} {Negative magnetoresistance without well-defined chirality in the weyl semimetal tap},\ }\href {https://doi.org/10.1038/ncomms11615} {\bibfield  {journal} {\bibinfo  {journal} {Nature Communications}\ }\textbf {\bibinfo {volume} {7}},\ \bibinfo {pages} {11615} (\bibinfo {year} {2016})}\BibitemShut {NoStop}%
\bibitem [{\citenamefont {Ong}\ and\ \citenamefont {Liang}(2021)}]{ong_experimental_2021}%
  \BibitemOpen
  \bibfield  {author} {\bibinfo {author} {\bibfnamefont {N.~P.}\ \bibnamefont {Ong}}\ and\ \bibinfo {author} {\bibfnamefont {S.}~\bibnamefont {Liang}},\ }\bibfield  {title} {\bibinfo {title} {Experimental signatures of the chiral anomaly in {Dirac}–{Weyl} semimetals},\ }\href {https://doi.org/10.1038/s42254-021-00310-9} {\bibfield  {journal} {\bibinfo  {journal} {Nat Rev Phys}\ }\textbf {\bibinfo {volume} {3}},\ \bibinfo {pages} {394} (\bibinfo {year} {2021})},\ \bibinfo {note} {number: 6 Publisher: Nature Publishing Group}\BibitemShut {NoStop}%
\bibitem [{\citenamefont {Flicker}\ \emph {et~al.}(2018)\citenamefont {Flicker}, \citenamefont {de~Juan}, \citenamefont {Bradlyn}, \citenamefont {Morimoto}, \citenamefont {Vergniory},\ and\ \citenamefont {Grushin}}]{Flicker2018}%
  \BibitemOpen
  \bibfield  {author} {\bibinfo {author} {\bibfnamefont {F.}~\bibnamefont {Flicker}}, \bibinfo {author} {\bibfnamefont {F.}~\bibnamefont {de~Juan}}, \bibinfo {author} {\bibfnamefont {B.}~\bibnamefont {Bradlyn}}, \bibinfo {author} {\bibfnamefont {T.}~\bibnamefont {Morimoto}}, \bibinfo {author} {\bibfnamefont {M.~G.}\ \bibnamefont {Vergniory}},\ and\ \bibinfo {author} {\bibfnamefont {A.~G.}\ \bibnamefont {Grushin}},\ }\bibfield  {title} {\bibinfo {title} {Chiral optical response of multifold fermions},\ }\href {https://doi.org/10.1103/PhysRevB.98.155145} {\bibfield  {journal} {\bibinfo  {journal} {Phys. Rev. B}\ }\textbf {\bibinfo {volume} {98}},\ \bibinfo {pages} {155145} (\bibinfo {year} {2018})}\BibitemShut {NoStop}%
\bibitem [{\citenamefont {Xiao}\ \emph {et~al.}(2010)\citenamefont {Xiao}, \citenamefont {Chang},\ and\ \citenamefont {Niu}}]{Xiao2010}%
  \BibitemOpen
  \bibfield  {author} {\bibinfo {author} {\bibfnamefont {D.}~\bibnamefont {Xiao}}, \bibinfo {author} {\bibfnamefont {M.-C.}\ \bibnamefont {Chang}},\ and\ \bibinfo {author} {\bibfnamefont {Q.}~\bibnamefont {Niu}},\ }\bibfield  {title} {\bibinfo {title} {{Berry phase effects on electronic properties}},\ }\href {https://doi.org/10.1103/revmodphys.82.1959} {\bibfield  {journal} {\bibinfo  {journal} {Reviews of Modern Physics}\ }\textbf {\bibinfo {volume} {82}},\ \bibinfo {pages} {1959 } (\bibinfo {year} {2010})}\BibitemShut {NoStop}%
\bibitem [{\citenamefont {Morimoto}\ \emph {et~al.}(2016)\citenamefont {Morimoto}, \citenamefont {Zhong}, \citenamefont {Orenstein},\ and\ \citenamefont {Moore}}]{Morimoto2016}%
  \BibitemOpen
  \bibfield  {author} {\bibinfo {author} {\bibfnamefont {T.}~\bibnamefont {Morimoto}}, \bibinfo {author} {\bibfnamefont {S.}~\bibnamefont {Zhong}}, \bibinfo {author} {\bibfnamefont {J.}~\bibnamefont {Orenstein}},\ and\ \bibinfo {author} {\bibfnamefont {J.~E.}\ \bibnamefont {Moore}},\ }\bibfield  {title} {\bibinfo {title} {Semiclassical theory of nonlinear magneto-optical responses with applications to topological dirac/weyl semimetals},\ }\href {https://doi.org/10.1103/PhysRevB.94.245121} {\bibfield  {journal} {\bibinfo  {journal} {Phys. Rev. B}\ }\textbf {\bibinfo {volume} {94}},\ \bibinfo {pages} {245121} (\bibinfo {year} {2016})}\BibitemShut {NoStop}%
\bibitem [{\citenamefont {Liang}\ \emph {et~al.}(2018{\natexlab{a}})\citenamefont {Liang}, \citenamefont {Lin}, \citenamefont {Kushwaha}, \citenamefont {Xing}, \citenamefont {Ni}, \citenamefont {Cava},\ and\ \citenamefont {Ong}}]{Liang:2018cj}%
  \BibitemOpen
  \bibfield  {author} {\bibinfo {author} {\bibfnamefont {S.}~\bibnamefont {Liang}}, \bibinfo {author} {\bibfnamefont {J.}~\bibnamefont {Lin}}, \bibinfo {author} {\bibfnamefont {S.}~\bibnamefont {Kushwaha}}, \bibinfo {author} {\bibfnamefont {J.}~\bibnamefont {Xing}}, \bibinfo {author} {\bibfnamefont {N.}~\bibnamefont {Ni}}, \bibinfo {author} {\bibfnamefont {R.~J.}\ \bibnamefont {Cava}},\ and\ \bibinfo {author} {\bibfnamefont {N.~P.}\ \bibnamefont {Ong}},\ }\bibfield  {title} {\bibinfo {title} {{Experimental Tests of the Chiral Anomaly Magnetoresistance in the Dirac-Weyl Semimetals Na3Bi and GdPtBi}},\ }\href {https://doi.org/10.1103/physrevx.8.031002} {\bibfield  {journal} {\bibinfo  {journal} {Physical Review X}\ }\textbf {\bibinfo {volume} {8}},\ \bibinfo {pages} {031002} (\bibinfo {year} {2018}{\natexlab{a}})}\BibitemShut {NoStop}%
\bibitem [{\citenamefont {Takane}\ \emph {et~al.}(2019)\citenamefont {Takane}, \citenamefont {Wang}, \citenamefont {Souma}, \citenamefont {Nakayama}, \citenamefont {Nakamura}, \citenamefont {Oinuma}, \citenamefont {Nakata}, \citenamefont {Iwasawa}, \citenamefont {Cacho}, \citenamefont {Kim}, \citenamefont {Horiba}, \citenamefont {Kumigashira}, \citenamefont {Takahashi}, \citenamefont {Ando},\ and\ \citenamefont {Sato}}]{takane_observation_2019}%
  \BibitemOpen
  \bibfield  {author} {\bibinfo {author} {\bibfnamefont {D.}~\bibnamefont {Takane}}, \bibinfo {author} {\bibfnamefont {Z.}~\bibnamefont {Wang}}, \bibinfo {author} {\bibfnamefont {S.}~\bibnamefont {Souma}}, \bibinfo {author} {\bibfnamefont {K.}~\bibnamefont {Nakayama}}, \bibinfo {author} {\bibfnamefont {T.}~\bibnamefont {Nakamura}}, \bibinfo {author} {\bibfnamefont {H.}~\bibnamefont {Oinuma}}, \bibinfo {author} {\bibfnamefont {Y.}~\bibnamefont {Nakata}}, \bibinfo {author} {\bibfnamefont {H.}~\bibnamefont {Iwasawa}}, \bibinfo {author} {\bibfnamefont {C.}~\bibnamefont {Cacho}}, \bibinfo {author} {\bibfnamefont {T.}~\bibnamefont {Kim}}, \bibinfo {author} {\bibfnamefont {K.}~\bibnamefont {Horiba}}, \bibinfo {author} {\bibfnamefont {H.}~\bibnamefont {Kumigashira}}, \bibinfo {author} {\bibfnamefont {T.}~\bibnamefont {Takahashi}}, \bibinfo {author} {\bibfnamefont {Y.}~\bibnamefont {Ando}},\ and\ \bibinfo {author} {\bibfnamefont {T.}~\bibnamefont {Sato}},\ }\bibfield  {title} {\bibinfo {title} {Observation of {Chiral}
  {Fermions} with a {Large} {Topological} {Charge} and and {Associated} {Fermi}-{Arc} {Surface} {States} in {CoSi}},\ }\href {https://doi.org/10.1103/PhysRevLett.122.076402} {\bibfield  {journal} {\bibinfo  {journal} {Phys. Rev. Lett.}\ }\textbf {\bibinfo {volume} {122}},\ \bibinfo {pages} {076402} (\bibinfo {year} {2019})},\ \bibinfo {note} {arXiv:1809.01312 [cond-mat]}\BibitemShut {NoStop}%
\bibitem [{\citenamefont {Rao}\ \emph {et~al.}(2019)\citenamefont {Rao}, \citenamefont {Li}, \citenamefont {Zhang}, \citenamefont {Tian}, \citenamefont {Li}, \citenamefont {Fu}, \citenamefont {Tang}, \citenamefont {Wang}, \citenamefont {Li}, \citenamefont {Fan} \emph {et~al.}}]{raoNature2019}%
  \BibitemOpen
  \bibfield  {author} {\bibinfo {author} {\bibfnamefont {Z.}~\bibnamefont {Rao}}, \bibinfo {author} {\bibfnamefont {H.}~\bibnamefont {Li}}, \bibinfo {author} {\bibfnamefont {T.}~\bibnamefont {Zhang}}, \bibinfo {author} {\bibfnamefont {S.}~\bibnamefont {Tian}}, \bibinfo {author} {\bibfnamefont {C.}~\bibnamefont {Li}}, \bibinfo {author} {\bibfnamefont {B.}~\bibnamefont {Fu}}, \bibinfo {author} {\bibfnamefont {C.}~\bibnamefont {Tang}}, \bibinfo {author} {\bibfnamefont {L.}~\bibnamefont {Wang}}, \bibinfo {author} {\bibfnamefont {Z.}~\bibnamefont {Li}}, \bibinfo {author} {\bibfnamefont {W.}~\bibnamefont {Fan}}, \emph {et~al.},\ }\bibfield  {title} {\bibinfo {title} {Observation of unconventional chiral fermions with long {Fermi} arcs in {CoSi}},\ }\href {https://www.nature.com/articles/s41586-019-1031-8} {\bibfield  {journal} {\bibinfo  {journal} {Nature}\ }\textbf {\bibinfo {volume} {567}},\ \bibinfo {pages} {496} (\bibinfo {year} {2019})}\BibitemShut {NoStop}%
\bibitem [{\citenamefont {Sanchez}\ \emph {et~al.}(2019)\citenamefont {Sanchez}, \citenamefont {Belopolski}, \citenamefont {Cochran}, \citenamefont {Xu}, \citenamefont {Yin}, \citenamefont {Chang}, \citenamefont {Xie}, \citenamefont {Manna}, \citenamefont {Süß}, \citenamefont {Huang}, \citenamefont {Alidoust}, \citenamefont {Multer}, \citenamefont {Zhang}, \citenamefont {Shumiya}, \citenamefont {Wang}, \citenamefont {Wang}, \citenamefont {Chang}, \citenamefont {Felser}, \citenamefont {Xu}, \citenamefont {Jia}, \citenamefont {Lin},\ and\ \citenamefont {Hasan}}]{sanchez_topological_2019}%
  \BibitemOpen
  \bibfield  {author} {\bibinfo {author} {\bibfnamefont {D.~S.}\ \bibnamefont {Sanchez}}, \bibinfo {author} {\bibfnamefont {I.}~\bibnamefont {Belopolski}}, \bibinfo {author} {\bibfnamefont {T.~A.}\ \bibnamefont {Cochran}}, \bibinfo {author} {\bibfnamefont {X.}~\bibnamefont {Xu}}, \bibinfo {author} {\bibfnamefont {J.-X.}\ \bibnamefont {Yin}}, \bibinfo {author} {\bibfnamefont {G.}~\bibnamefont {Chang}}, \bibinfo {author} {\bibfnamefont {W.}~\bibnamefont {Xie}}, \bibinfo {author} {\bibfnamefont {K.}~\bibnamefont {Manna}}, \bibinfo {author} {\bibfnamefont {V.}~\bibnamefont {Süß}}, \bibinfo {author} {\bibfnamefont {C.-Y.}\ \bibnamefont {Huang}}, \bibinfo {author} {\bibfnamefont {N.}~\bibnamefont {Alidoust}}, \bibinfo {author} {\bibfnamefont {D.}~\bibnamefont {Multer}}, \bibinfo {author} {\bibfnamefont {S.~S.}\ \bibnamefont {Zhang}}, \bibinfo {author} {\bibfnamefont {N.}~\bibnamefont {Shumiya}}, \bibinfo {author} {\bibfnamefont {X.}~\bibnamefont {Wang}}, \bibinfo {author} {\bibfnamefont {G.-Q.}\ \bibnamefont
  {Wang}}, \bibinfo {author} {\bibfnamefont {T.-R.}\ \bibnamefont {Chang}}, \bibinfo {author} {\bibfnamefont {C.}~\bibnamefont {Felser}}, \bibinfo {author} {\bibfnamefont {S.-Y.}\ \bibnamefont {Xu}}, \bibinfo {author} {\bibfnamefont {S.}~\bibnamefont {Jia}}, \bibinfo {author} {\bibfnamefont {H.}~\bibnamefont {Lin}},\ and\ \bibinfo {author} {\bibfnamefont {M.~Z.}\ \bibnamefont {Hasan}},\ }\bibfield  {title} {\bibinfo {title} {Topological chiral crystals with helicoid-arc quantum states},\ }\href {https://doi.org/10.1038/s41586-019-1037-2} {\bibfield  {journal} {\bibinfo  {journal} {Nature}\ }\textbf {\bibinfo {volume} {567}},\ \bibinfo {pages} {500} (\bibinfo {year} {2019})},\ \bibinfo {note} {bandiera\_abtest: a Cg\_type: Nature Research Journals Number: 7749 Primary\_atype: Research Publisher: Nature Publishing Group Subject\_term: Electronic properties and materials;Topological matter Subject\_term\_id: electronic-properties-and-materials;topological-matter}\BibitemShut {NoStop}%
\bibitem [{\citenamefont {Schr{\"o}ter}\ \emph {et~al.}(2019)\citenamefont {Schr{\"o}ter}, \citenamefont {Pei}, \citenamefont {Vergniory}, \citenamefont {Sun}, \citenamefont {Manna}, \citenamefont {de~Juan}, \citenamefont {Krieger}, \citenamefont {S{\"u}ss}, \citenamefont {Schmidt}, \citenamefont {Dudin}, \citenamefont {Bradlyn}, \citenamefont {Kim}, \citenamefont {Schmitt}, \citenamefont {Cacho}, \citenamefont {Felser}, \citenamefont {Strocov},\ and\ \citenamefont {Chen}}]{Schroeter2019}%
  \BibitemOpen
  \bibfield  {author} {\bibinfo {author} {\bibfnamefont {N.~B.~M.}\ \bibnamefont {Schr{\"o}ter}}, \bibinfo {author} {\bibfnamefont {D.}~\bibnamefont {Pei}}, \bibinfo {author} {\bibfnamefont {M.~G.}\ \bibnamefont {Vergniory}}, \bibinfo {author} {\bibfnamefont {Y.}~\bibnamefont {Sun}}, \bibinfo {author} {\bibfnamefont {K.}~\bibnamefont {Manna}}, \bibinfo {author} {\bibfnamefont {F.}~\bibnamefont {de~Juan}}, \bibinfo {author} {\bibfnamefont {J.~A.}\ \bibnamefont {Krieger}}, \bibinfo {author} {\bibfnamefont {V.}~\bibnamefont {S{\"u}ss}}, \bibinfo {author} {\bibfnamefont {M.}~\bibnamefont {Schmidt}}, \bibinfo {author} {\bibfnamefont {P.}~\bibnamefont {Dudin}}, \bibinfo {author} {\bibfnamefont {B.}~\bibnamefont {Bradlyn}}, \bibinfo {author} {\bibfnamefont {T.~K.}\ \bibnamefont {Kim}}, \bibinfo {author} {\bibfnamefont {T.}~\bibnamefont {Schmitt}}, \bibinfo {author} {\bibfnamefont {C.}~\bibnamefont {Cacho}}, \bibinfo {author} {\bibfnamefont {C.}~\bibnamefont {Felser}}, \bibinfo {author} {\bibfnamefont {V.~N.}\
  \bibnamefont {Strocov}},\ and\ \bibinfo {author} {\bibfnamefont {Y.}~\bibnamefont {Chen}},\ }\bibfield  {title} {\bibinfo {title} {Chiral topological semimetal with multifold band crossings and long fermi arcs},\ }\href {https://doi.org/10.1038/s41567-019-0511-y} {\bibfield  {journal} {\bibinfo  {journal} {Nature Physics}\ }\textbf {\bibinfo {volume} {15}},\ \bibinfo {pages} {759} (\bibinfo {year} {2019})}\BibitemShut {NoStop}%
\bibitem [{\citenamefont {Schr{\"o}ter}\ \emph {et~al.}(2020)\citenamefont {Schr{\"o}ter}, \citenamefont {Stolz}, \citenamefont {Manna}, \citenamefont {de~Juan}, \citenamefont {Vergniory}, \citenamefont {Krieger}, \citenamefont {Pei}, \citenamefont {Schmitt}, \citenamefont {Dudin}, \citenamefont {Kim}, \citenamefont {Cacho}, \citenamefont {Bradlyn}, \citenamefont {Borrmann}, \citenamefont {Schmidt}, \citenamefont {Widmer}, \citenamefont {Strocov},\ and\ \citenamefont {Felser}}]{Schroter179}%
  \BibitemOpen
  \bibfield  {author} {\bibinfo {author} {\bibfnamefont {N.~B.~M.}\ \bibnamefont {Schr{\"o}ter}}, \bibinfo {author} {\bibfnamefont {S.}~\bibnamefont {Stolz}}, \bibinfo {author} {\bibfnamefont {K.}~\bibnamefont {Manna}}, \bibinfo {author} {\bibfnamefont {F.}~\bibnamefont {de~Juan}}, \bibinfo {author} {\bibfnamefont {M.~G.}\ \bibnamefont {Vergniory}}, \bibinfo {author} {\bibfnamefont {J.~A.}\ \bibnamefont {Krieger}}, \bibinfo {author} {\bibfnamefont {D.}~\bibnamefont {Pei}}, \bibinfo {author} {\bibfnamefont {T.}~\bibnamefont {Schmitt}}, \bibinfo {author} {\bibfnamefont {P.}~\bibnamefont {Dudin}}, \bibinfo {author} {\bibfnamefont {T.~K.}\ \bibnamefont {Kim}}, \bibinfo {author} {\bibfnamefont {C.}~\bibnamefont {Cacho}}, \bibinfo {author} {\bibfnamefont {B.}~\bibnamefont {Bradlyn}}, \bibinfo {author} {\bibfnamefont {H.}~\bibnamefont {Borrmann}}, \bibinfo {author} {\bibfnamefont {M.}~\bibnamefont {Schmidt}}, \bibinfo {author} {\bibfnamefont {R.}~\bibnamefont {Widmer}}, \bibinfo {author} {\bibfnamefont {V.~N.}\
  \bibnamefont {Strocov}},\ and\ \bibinfo {author} {\bibfnamefont {C.}~\bibnamefont {Felser}},\ }\bibfield  {title} {\bibinfo {title} {Observation and control of maximal chern numbers in a chiral topological semimetal},\ }\href {https://doi.org/10.1126/science.aaz3480} {\bibfield  {journal} {\bibinfo  {journal} {Science}\ }\textbf {\bibinfo {volume} {369}},\ \bibinfo {pages} {179} (\bibinfo {year} {2020})}\BibitemShut {NoStop}%
\bibitem [{\citenamefont {Yao}\ \emph {et~al.}(2020)\citenamefont {Yao}, \citenamefont {Manna}, \citenamefont {Yang}, \citenamefont {Fedorov}, \citenamefont {Voroshnin}, \citenamefont {Schwarze}, \citenamefont {Hornung}, \citenamefont {Chattopadhyay}, \citenamefont {Sun}, \citenamefont {Guin}, \citenamefont {Wosnitza}, \citenamefont {Borrmann}, \citenamefont {Shekhar}, \citenamefont {Kumar}, \citenamefont {Fink}, \citenamefont {Sun},\ and\ \citenamefont {Felser}}]{Yao:2020cc}%
  \BibitemOpen
  \bibfield  {author} {\bibinfo {author} {\bibfnamefont {M.}~\bibnamefont {Yao}}, \bibinfo {author} {\bibfnamefont {K.}~\bibnamefont {Manna}}, \bibinfo {author} {\bibfnamefont {Q.}~\bibnamefont {Yang}}, \bibinfo {author} {\bibfnamefont {A.}~\bibnamefont {Fedorov}}, \bibinfo {author} {\bibfnamefont {V.}~\bibnamefont {Voroshnin}}, \bibinfo {author} {\bibfnamefont {B.~V.}\ \bibnamefont {Schwarze}}, \bibinfo {author} {\bibfnamefont {J.}~\bibnamefont {Hornung}}, \bibinfo {author} {\bibfnamefont {S.}~\bibnamefont {Chattopadhyay}}, \bibinfo {author} {\bibfnamefont {Z.}~\bibnamefont {Sun}}, \bibinfo {author} {\bibfnamefont {S.~N.}\ \bibnamefont {Guin}}, \bibinfo {author} {\bibfnamefont {J.}~\bibnamefont {Wosnitza}}, \bibinfo {author} {\bibfnamefont {H.}~\bibnamefont {Borrmann}}, \bibinfo {author} {\bibfnamefont {C.}~\bibnamefont {Shekhar}}, \bibinfo {author} {\bibfnamefont {N.}~\bibnamefont {Kumar}}, \bibinfo {author} {\bibfnamefont {J.}~\bibnamefont {Fink}}, \bibinfo {author} {\bibfnamefont {Y.}~\bibnamefont
  {Sun}},\ and\ \bibinfo {author} {\bibfnamefont {C.}~\bibnamefont {Felser}},\ }\bibfield  {title} {\bibinfo {title} {{Observation of giant spin-split Fermi-arc with maximal Chern number in the chiral topological semimetal PtGa}},\ }\href {https://www.nature.com/articles/s41467-020-15865-x} {\bibfield  {journal} {\bibinfo  {journal} {Nature Communications}\ }\textbf {\bibinfo {volume} {11}},\ \bibinfo {pages} {1} (\bibinfo {year} {2020})}\BibitemShut {NoStop}%
\bibitem [{\citenamefont {Sessi}\ \emph {et~al.}(2020)\citenamefont {Sessi}, \citenamefont {Fan}, \citenamefont {K{\"u}ster}, \citenamefont {Manna}, \citenamefont {Schr{\"o}ter}, \citenamefont {Ji}, \citenamefont {Stolz}, \citenamefont {Krieger}, \citenamefont {Pei}, \citenamefont {Kim}, \citenamefont {Dudin}, \citenamefont {Cacho}, \citenamefont {Widmer}, \citenamefont {Borrmann}, \citenamefont {Shi}, \citenamefont {Chang}, \citenamefont {Sun}, \citenamefont {Felser},\ and\ \citenamefont {Parkin}}]{Sessi:2020dw}%
  \BibitemOpen
  \bibfield  {author} {\bibinfo {author} {\bibfnamefont {P.}~\bibnamefont {Sessi}}, \bibinfo {author} {\bibfnamefont {F.-R.}\ \bibnamefont {Fan}}, \bibinfo {author} {\bibfnamefont {F.}~\bibnamefont {K{\"u}ster}}, \bibinfo {author} {\bibfnamefont {K.}~\bibnamefont {Manna}}, \bibinfo {author} {\bibfnamefont {N.~B.~M.}\ \bibnamefont {Schr{\"o}ter}}, \bibinfo {author} {\bibfnamefont {J.-R.}\ \bibnamefont {Ji}}, \bibinfo {author} {\bibfnamefont {S.}~\bibnamefont {Stolz}}, \bibinfo {author} {\bibfnamefont {J.~A.}\ \bibnamefont {Krieger}}, \bibinfo {author} {\bibfnamefont {D.}~\bibnamefont {Pei}}, \bibinfo {author} {\bibfnamefont {T.~K.}\ \bibnamefont {Kim}}, \bibinfo {author} {\bibfnamefont {P.}~\bibnamefont {Dudin}}, \bibinfo {author} {\bibfnamefont {C.}~\bibnamefont {Cacho}}, \bibinfo {author} {\bibfnamefont {R.}~\bibnamefont {Widmer}}, \bibinfo {author} {\bibfnamefont {H.}~\bibnamefont {Borrmann}}, \bibinfo {author} {\bibfnamefont {W.}~\bibnamefont {Shi}}, \bibinfo {author} {\bibfnamefont {K.}~\bibnamefont
  {Chang}}, \bibinfo {author} {\bibfnamefont {Y.}~\bibnamefont {Sun}}, \bibinfo {author} {\bibfnamefont {C.}~\bibnamefont {Felser}},\ and\ \bibinfo {author} {\bibfnamefont {S.~S.~P.}\ \bibnamefont {Parkin}},\ }\bibfield  {title} {\bibinfo {title} {{Handedness-dependent quasiparticle interference in the two enantiomers of the topological chiral semimetal PdGa}},\ }\href {https://www.nature.com/articles/s41467-020-17261-x} {\bibfield  {journal} {\bibinfo  {journal} {Nature Communications}\ }\textbf {\bibinfo {volume} {11}},\ \bibinfo {pages} {1} (\bibinfo {year} {2020})}\BibitemShut {NoStop}%
\bibitem [{\citenamefont {Chang}\ \emph {et~al.}(2017)\citenamefont {Chang}, \citenamefont {Xu}, \citenamefont {Wieder}, \citenamefont {Sanchez}, \citenamefont {Huang}, \citenamefont {Belopolski}, \citenamefont {Chang}, \citenamefont {Zhang}, \citenamefont {Bansil}, \citenamefont {Lin},\ and\ \citenamefont {Hasan}}]{chang_unconventional_2017}%
  \BibitemOpen
  \bibfield  {author} {\bibinfo {author} {\bibfnamefont {G.}~\bibnamefont {Chang}}, \bibinfo {author} {\bibfnamefont {S.-Y.}\ \bibnamefont {Xu}}, \bibinfo {author} {\bibfnamefont {B.~J.}\ \bibnamefont {Wieder}}, \bibinfo {author} {\bibfnamefont {D.~S.}\ \bibnamefont {Sanchez}}, \bibinfo {author} {\bibfnamefont {S.-M.}\ \bibnamefont {Huang}}, \bibinfo {author} {\bibfnamefont {I.}~\bibnamefont {Belopolski}}, \bibinfo {author} {\bibfnamefont {T.-R.}\ \bibnamefont {Chang}}, \bibinfo {author} {\bibfnamefont {S.}~\bibnamefont {Zhang}}, \bibinfo {author} {\bibfnamefont {A.}~\bibnamefont {Bansil}}, \bibinfo {author} {\bibfnamefont {H.}~\bibnamefont {Lin}},\ and\ \bibinfo {author} {\bibfnamefont {M.~Z.}\ \bibnamefont {Hasan}},\ }\bibfield  {title} {\bibinfo {title} {Unconventional {Chiral} {Fermions} and {Large} {Topological} {Fermi} {Arcs} in {RhSi}},\ }\href {https://doi.org/10.1103/PhysRevLett.119.206401} {\bibfield  {journal} {\bibinfo  {journal} {Phys. Rev. Lett.}\ }\textbf {\bibinfo {volume} {119}},\ \bibinfo
  {pages} {206401} (\bibinfo {year} {2017})},\ \bibinfo {note} {publisher: American Physical Society}\BibitemShut {NoStop}%
\bibitem [{\citenamefont {Xu}\ \emph {et~al.}(2020)\citenamefont {Xu}, \citenamefont {Fang}, \citenamefont {Sanchez-Martinez}, \citenamefont {Venderbos}, \citenamefont {Ni}, \citenamefont {Qiu}, \citenamefont {Manna}, \citenamefont {Wang}, \citenamefont {Paglione}, \citenamefont {Bernhard}, \citenamefont {Felser}, \citenamefont {Mele}, \citenamefont {Grushin}, \citenamefont {Rappe},\ and\ \citenamefont {Wu}}]{Xu:2020ei}%
  \BibitemOpen
  \bibfield  {author} {\bibinfo {author} {\bibfnamefont {B.}~\bibnamefont {Xu}}, \bibinfo {author} {\bibfnamefont {Z.}~\bibnamefont {Fang}}, \bibinfo {author} {\bibfnamefont {M.-A.}\ \bibnamefont {Sanchez-Martinez}}, \bibinfo {author} {\bibfnamefont {J.~W.~F.}\ \bibnamefont {Venderbos}}, \bibinfo {author} {\bibfnamefont {Z.}~\bibnamefont {Ni}}, \bibinfo {author} {\bibfnamefont {T.}~\bibnamefont {Qiu}}, \bibinfo {author} {\bibfnamefont {K.}~\bibnamefont {Manna}}, \bibinfo {author} {\bibfnamefont {K.}~\bibnamefont {Wang}}, \bibinfo {author} {\bibfnamefont {J.}~\bibnamefont {Paglione}}, \bibinfo {author} {\bibfnamefont {C.}~\bibnamefont {Bernhard}}, \bibinfo {author} {\bibfnamefont {C.}~\bibnamefont {Felser}}, \bibinfo {author} {\bibfnamefont {E.~J.}\ \bibnamefont {Mele}}, \bibinfo {author} {\bibfnamefont {A.~G.}\ \bibnamefont {Grushin}}, \bibinfo {author} {\bibfnamefont {A.~M.}\ \bibnamefont {Rappe}},\ and\ \bibinfo {author} {\bibfnamefont {L.}~\bibnamefont {Wu}},\ }\bibfield  {title} {\bibinfo {title}
  {{Optical signatures of multifold fermions in the chiral topological semimetal CoSi}},\ }\href {https://doi.org/10.1073/pnas.2010752117} {\bibfield  {journal} {\bibinfo  {journal} {Proceedings of the National Academy of Sciences}\ }\textbf {\bibinfo {volume} {83}},\ \bibinfo {pages} {202010752 } (\bibinfo {year} {2020})}\BibitemShut {NoStop}%
\bibitem [{\citenamefont {Ni}\ \emph {et~al.}(2021)\citenamefont {Ni}, \citenamefont {Wang}, \citenamefont {Zhang}, \citenamefont {Pozo}, \citenamefont {Xu}, \citenamefont {Han}, \citenamefont {Manna}, \citenamefont {Paglione}, \citenamefont {Felser}, \citenamefont {Grushin}, \citenamefont {Juan}, \citenamefont {Mele},\ and\ \citenamefont {Wu}}]{Ni:2021cta}%
  \BibitemOpen
  \bibfield  {author} {\bibinfo {author} {\bibfnamefont {Z.}~\bibnamefont {Ni}}, \bibinfo {author} {\bibfnamefont {K.}~\bibnamefont {Wang}}, \bibinfo {author} {\bibfnamefont {Y.}~\bibnamefont {Zhang}}, \bibinfo {author} {\bibfnamefont {O.}~\bibnamefont {Pozo}}, \bibinfo {author} {\bibfnamefont {B.}~\bibnamefont {Xu}}, \bibinfo {author} {\bibfnamefont {X.}~\bibnamefont {Han}}, \bibinfo {author} {\bibfnamefont {K.}~\bibnamefont {Manna}}, \bibinfo {author} {\bibfnamefont {J.}~\bibnamefont {Paglione}}, \bibinfo {author} {\bibfnamefont {C.}~\bibnamefont {Felser}}, \bibinfo {author} {\bibfnamefont {A.~G.}\ \bibnamefont {Grushin}}, \bibinfo {author} {\bibfnamefont {F.~d.}\ \bibnamefont {Juan}}, \bibinfo {author} {\bibfnamefont {E.~J.}\ \bibnamefont {Mele}},\ and\ \bibinfo {author} {\bibfnamefont {L.}~\bibnamefont {Wu}},\ }\bibfield  {title} {\bibinfo {title} {{Giant topological longitudinal circular photo-galvanic effect in the chiral multifold semimetal CoSi}},\ }\href {https://doi.org/10.1038/s41467-020-20408-5}
  {\bibfield  {journal} {\bibinfo  {journal} {Nature Communications}\ }\textbf {\bibinfo {volume} {12}},\ \bibinfo {pages} {R935} (\bibinfo {year} {2021})}\BibitemShut {NoStop}%
\bibitem [{\citenamefont {Xu}\ \emph {et~al.}(2019)\citenamefont {Xu}, \citenamefont {Wang}, \citenamefont {Cochran}, \citenamefont {Sanchez}, \citenamefont {Chang}, \citenamefont {Belopolski}, \citenamefont {Wang}, \citenamefont {Liu}, \citenamefont {Tien}, \citenamefont {Gui}, \citenamefont {Xie}, \citenamefont {Hasan}, \citenamefont {Chang},\ and\ \citenamefont {Jia}}]{xu_crystal_2019}%
  \BibitemOpen
  \bibfield  {author} {\bibinfo {author} {\bibfnamefont {X.}~\bibnamefont {Xu}}, \bibinfo {author} {\bibfnamefont {X.}~\bibnamefont {Wang}}, \bibinfo {author} {\bibfnamefont {T.~A.}\ \bibnamefont {Cochran}}, \bibinfo {author} {\bibfnamefont {D.~S.}\ \bibnamefont {Sanchez}}, \bibinfo {author} {\bibfnamefont {G.}~\bibnamefont {Chang}}, \bibinfo {author} {\bibfnamefont {I.}~\bibnamefont {Belopolski}}, \bibinfo {author} {\bibfnamefont {G.}~\bibnamefont {Wang}}, \bibinfo {author} {\bibfnamefont {Y.}~\bibnamefont {Liu}}, \bibinfo {author} {\bibfnamefont {H.-J.}\ \bibnamefont {Tien}}, \bibinfo {author} {\bibfnamefont {X.}~\bibnamefont {Gui}}, \bibinfo {author} {\bibfnamefont {W.}~\bibnamefont {Xie}}, \bibinfo {author} {\bibfnamefont {M.~Z.}\ \bibnamefont {Hasan}}, \bibinfo {author} {\bibfnamefont {T.-R.}\ \bibnamefont {Chang}},\ and\ \bibinfo {author} {\bibfnamefont {S.}~\bibnamefont {Jia}},\ }\bibfield  {title} {\bibinfo {title} {Crystal growth and quantum oscillations in the topological chiral semimetal {CoSi}},\
  }\href {https://doi.org/10.1103/PhysRevB.100.045104} {\bibfield  {journal} {\bibinfo  {journal} {Phys. Rev. B}\ }\textbf {\bibinfo {volume} {100}},\ \bibinfo {pages} {045104} (\bibinfo {year} {2019})},\ \bibinfo {note} {publisher: American Physical Society}\BibitemShut {NoStop}%
\bibitem [{\citenamefont {Hirschberger}\ \emph {et~al.}(2016)\citenamefont {Hirschberger}, \citenamefont {Kushwaha}, \citenamefont {Wang}, \citenamefont {Gibson}, \citenamefont {Liang}, \citenamefont {Belvin}, \citenamefont {Bernevig}, \citenamefont {Cava},\ and\ \citenamefont {Ong}}]{hirschberger_chiral_2016}%
  \BibitemOpen
  \bibfield  {author} {\bibinfo {author} {\bibfnamefont {M.}~\bibnamefont {Hirschberger}}, \bibinfo {author} {\bibfnamefont {S.}~\bibnamefont {Kushwaha}}, \bibinfo {author} {\bibfnamefont {Z.}~\bibnamefont {Wang}}, \bibinfo {author} {\bibfnamefont {Q.}~\bibnamefont {Gibson}}, \bibinfo {author} {\bibfnamefont {S.}~\bibnamefont {Liang}}, \bibinfo {author} {\bibfnamefont {C.~A.}\ \bibnamefont {Belvin}}, \bibinfo {author} {\bibfnamefont {B.~A.}\ \bibnamefont {Bernevig}}, \bibinfo {author} {\bibfnamefont {R.~J.}\ \bibnamefont {Cava}},\ and\ \bibinfo {author} {\bibfnamefont {N.~P.}\ \bibnamefont {Ong}},\ }\bibfield  {title} {\bibinfo {title} {The chiral anomaly and thermopower of {Weyl} fermions in the half-{Heusler} {GdPtBi}},\ }\href {https://doi.org/10.1038/nmat4684} {\bibfield  {journal} {\bibinfo  {journal} {Nature Mater}\ }\textbf {\bibinfo {volume} {15}},\ \bibinfo {pages} {1161} (\bibinfo {year} {2016})},\ \bibinfo {note} {number: 11 Publisher: Nature Publishing Group}\BibitemShut {NoStop}%
\bibitem [{\citenamefont {Niemann}\ \emph {et~al.}(2017)\citenamefont {Niemann}, \citenamefont {Gooth}, \citenamefont {Wu}, \citenamefont {Bäßler}, \citenamefont {Sergelius}, \citenamefont {Hühne}, \citenamefont {Rellinghaus}, \citenamefont {Shekhar}, \citenamefont {Süß}, \citenamefont {Schmidt}, \citenamefont {Felser}, \citenamefont {Yan},\ and\ \citenamefont {Nielsch}}]{niemann_chiral_2017}%
  \BibitemOpen
  \bibfield  {author} {\bibinfo {author} {\bibfnamefont {A.~C.}\ \bibnamefont {Niemann}}, \bibinfo {author} {\bibfnamefont {J.}~\bibnamefont {Gooth}}, \bibinfo {author} {\bibfnamefont {S.-C.}\ \bibnamefont {Wu}}, \bibinfo {author} {\bibfnamefont {S.}~\bibnamefont {Bäßler}}, \bibinfo {author} {\bibfnamefont {P.}~\bibnamefont {Sergelius}}, \bibinfo {author} {\bibfnamefont {R.}~\bibnamefont {Hühne}}, \bibinfo {author} {\bibfnamefont {B.}~\bibnamefont {Rellinghaus}}, \bibinfo {author} {\bibfnamefont {C.}~\bibnamefont {Shekhar}}, \bibinfo {author} {\bibfnamefont {V.}~\bibnamefont {Süß}}, \bibinfo {author} {\bibfnamefont {M.}~\bibnamefont {Schmidt}}, \bibinfo {author} {\bibfnamefont {C.}~\bibnamefont {Felser}}, \bibinfo {author} {\bibfnamefont {B.}~\bibnamefont {Yan}},\ and\ \bibinfo {author} {\bibfnamefont {K.}~\bibnamefont {Nielsch}},\ }\bibfield  {title} {\bibinfo {title} {Chiral magnetoresistance in the {Weyl} semimetal {NbP}},\ }\href {https://doi.org/10.1038/srep43394} {\bibfield  {journal} {\bibinfo
  {journal} {Sci Rep}\ }\textbf {\bibinfo {volume} {7}},\ \bibinfo {pages} {43394} (\bibinfo {year} {2017})},\ \bibinfo {note} {number: 1 Publisher: Nature Publishing Group}\BibitemShut {NoStop}%
\bibitem [{\citenamefont {Xiong}\ \emph {et~al.}(2015)\citenamefont {Xiong}, \citenamefont {Kushwaha}, \citenamefont {Liang}, \citenamefont {Krizan}, \citenamefont {Hirschberger}, \citenamefont {Wang}, \citenamefont {Cava},\ and\ \citenamefont {Ong}}]{xiong_evidence_2015}%
  \BibitemOpen
  \bibfield  {author} {\bibinfo {author} {\bibfnamefont {J.}~\bibnamefont {Xiong}}, \bibinfo {author} {\bibfnamefont {S.~K.}\ \bibnamefont {Kushwaha}}, \bibinfo {author} {\bibfnamefont {T.}~\bibnamefont {Liang}}, \bibinfo {author} {\bibfnamefont {J.~W.}\ \bibnamefont {Krizan}}, \bibinfo {author} {\bibfnamefont {M.}~\bibnamefont {Hirschberger}}, \bibinfo {author} {\bibfnamefont {W.}~\bibnamefont {Wang}}, \bibinfo {author} {\bibfnamefont {R.~J.}\ \bibnamefont {Cava}},\ and\ \bibinfo {author} {\bibfnamefont {N.~P.}\ \bibnamefont {Ong}},\ }\bibfield  {title} {\bibinfo {title} {Evidence for the chiral anomaly in the {Dirac} semimetal {Na3Bi}},\ }\href {https://doi.org/10.1126/science.aac6089} {\bibfield  {journal} {\bibinfo  {journal} {Science}\ }\textbf {\bibinfo {volume} {350}},\ \bibinfo {pages} {413} (\bibinfo {year} {2015})},\ \bibinfo {note} {publisher: American Association for the Advancement of Science}\BibitemShut {NoStop}%
\bibitem [{\citenamefont {Li}\ \emph {et~al.}(2016)\citenamefont {Li}, \citenamefont {He}, \citenamefont {Lu}, \citenamefont {Zhang}, \citenamefont {Liu}, \citenamefont {Ma}, \citenamefont {Fan}, \citenamefont {Shen},\ and\ \citenamefont {Wang}}]{li_negative_2016}%
  \BibitemOpen
  \bibfield  {author} {\bibinfo {author} {\bibfnamefont {H.}~\bibnamefont {Li}}, \bibinfo {author} {\bibfnamefont {H.}~\bibnamefont {He}}, \bibinfo {author} {\bibfnamefont {H.-Z.}\ \bibnamefont {Lu}}, \bibinfo {author} {\bibfnamefont {H.}~\bibnamefont {Zhang}}, \bibinfo {author} {\bibfnamefont {H.}~\bibnamefont {Liu}}, \bibinfo {author} {\bibfnamefont {R.}~\bibnamefont {Ma}}, \bibinfo {author} {\bibfnamefont {Z.}~\bibnamefont {Fan}}, \bibinfo {author} {\bibfnamefont {S.-Q.}\ \bibnamefont {Shen}},\ and\ \bibinfo {author} {\bibfnamefont {J.}~\bibnamefont {Wang}},\ }\bibfield  {title} {\bibinfo {title} {Negative magnetoresistance in {Dirac} semimetal {Cd3As2}},\ }\href {https://doi.org/10.1038/ncomms10301} {\bibfield  {journal} {\bibinfo  {journal} {Nat Commun}\ }\textbf {\bibinfo {volume} {7}},\ \bibinfo {pages} {10301} (\bibinfo {year} {2016})},\ \bibinfo {note} {number: 1 Publisher: Nature Publishing Group}\BibitemShut {NoStop}%
\bibitem [{\citenamefont {Yang}\ \emph {et~al.}(2015)\citenamefont {Yang}, \citenamefont {Liu}, \citenamefont {Wang}, \citenamefont {Zheng},\ and\ \citenamefont {Xu}}]{yang_chiral_2015}%
  \BibitemOpen
  \bibfield  {author} {\bibinfo {author} {\bibfnamefont {X.}~\bibnamefont {Yang}}, \bibinfo {author} {\bibfnamefont {Y.}~\bibnamefont {Liu}}, \bibinfo {author} {\bibfnamefont {Z.}~\bibnamefont {Wang}}, \bibinfo {author} {\bibfnamefont {Y.}~\bibnamefont {Zheng}},\ and\ \bibinfo {author} {\bibfnamefont {Z.-a.}\ \bibnamefont {Xu}},\ }\href {https://doi.org/10.48550/arXiv.1506.03190} {\bibinfo {title} {Chiral anomaly induced negative magnetoresistance in topological {Weyl} semimetal {NbAs}}} (\bibinfo {year} {2015}),\ \bibinfo {note} {arXiv:1506.03190 [cond-mat]}\BibitemShut {NoStop}%
\bibitem [{\citenamefont {Wang}\ \emph {et~al.}(2020)\citenamefont {Wang}, \citenamefont {Xu}, \citenamefont {Lu}, \citenamefont {Wang}, \citenamefont {Zeng}, \citenamefont {Lin}, \citenamefont {Liu}, \citenamefont {Lu},\ and\ \citenamefont {Xia}}]{wang_haas--van_2020}%
  \BibitemOpen
  \bibfield  {author} {\bibinfo {author} {\bibfnamefont {H.}~\bibnamefont {Wang}}, \bibinfo {author} {\bibfnamefont {S.}~\bibnamefont {Xu}}, \bibinfo {author} {\bibfnamefont {X.-Q.}\ \bibnamefont {Lu}}, \bibinfo {author} {\bibfnamefont {X.-Y.}\ \bibnamefont {Wang}}, \bibinfo {author} {\bibfnamefont {X.-Y.}\ \bibnamefont {Zeng}}, \bibinfo {author} {\bibfnamefont {J.-F.}\ \bibnamefont {Lin}}, \bibinfo {author} {\bibfnamefont {K.}~\bibnamefont {Liu}}, \bibinfo {author} {\bibfnamefont {Z.-Y.}\ \bibnamefont {Lu}},\ and\ \bibinfo {author} {\bibfnamefont {T.-L.}\ \bibnamefont {Xia}},\ }\bibfield  {title} {\bibinfo {title} {de {Haas}--van {Alphen} quantum oscillations and electronic structure in the large-{Chern}-number topological chiral semimetal {CoSi}},\ }\href {https://doi.org/10.1103/PhysRevB.102.115129} {\bibfield  {journal} {\bibinfo  {journal} {Phys. Rev. B}\ }\textbf {\bibinfo {volume} {102}},\ \bibinfo {pages} {115129} (\bibinfo {year} {2020})},\ \bibinfo {note} {publisher: American Physical
  Society}\BibitemShut {NoStop}%
\bibitem [{\citenamefont {Guo}\ \emph {et~al.}(2021)\citenamefont {Guo}, \citenamefont {Hu}, \citenamefont {Putzke}, \citenamefont {Diaz}, \citenamefont {Huang}, \citenamefont {Manna}, \citenamefont {Fan}, \citenamefont {Shekhar}, \citenamefont {Sun}, \citenamefont {Felser}, \citenamefont {Liu}, \citenamefont {Bernevig},\ and\ \citenamefont {Moll}}]{guo_hidden_2021}%
  \BibitemOpen
  \bibfield  {author} {\bibinfo {author} {\bibfnamefont {C.}~\bibnamefont {Guo}}, \bibinfo {author} {\bibfnamefont {L.}~\bibnamefont {Hu}}, \bibinfo {author} {\bibfnamefont {C.}~\bibnamefont {Putzke}}, \bibinfo {author} {\bibfnamefont {J.}~\bibnamefont {Diaz}}, \bibinfo {author} {\bibfnamefont {X.}~\bibnamefont {Huang}}, \bibinfo {author} {\bibfnamefont {K.}~\bibnamefont {Manna}}, \bibinfo {author} {\bibfnamefont {F.-R.}\ \bibnamefont {Fan}}, \bibinfo {author} {\bibfnamefont {C.}~\bibnamefont {Shekhar}}, \bibinfo {author} {\bibfnamefont {Y.}~\bibnamefont {Sun}}, \bibinfo {author} {\bibfnamefont {C.}~\bibnamefont {Felser}}, \bibinfo {author} {\bibfnamefont {C.}~\bibnamefont {Liu}}, \bibinfo {author} {\bibfnamefont {B.~A.}\ \bibnamefont {Bernevig}},\ and\ \bibinfo {author} {\bibfnamefont {P.~J.~W.}\ \bibnamefont {Moll}},\ }\bibfield  {title} {\bibinfo {title} {Hidden quasi-symmetries stabilize non-trivial quantum oscillations in {CoSi}},\ }\href {http://arxiv.org/abs/2108.02279} {\bibfield  {journal} {\bibinfo
  {journal} {arXiv:2108.02279 [cond-mat]}\ } (\bibinfo {year} {2021})},\ \bibinfo {note} {arXiv: 2108.02279}\BibitemShut {NoStop}%
\bibitem [{\citenamefont {Sasmal}\ \emph {et~al.}(2022)\citenamefont {Sasmal}, \citenamefont {Dwari}, \citenamefont {Maity}, \citenamefont {Saini}, \citenamefont {Thamizhavel},\ and\ \citenamefont {Mondal}}]{sasmal_shubnikov-haas_2022}%
  \BibitemOpen
  \bibfield  {author} {\bibinfo {author} {\bibfnamefont {S.}~\bibnamefont {Sasmal}}, \bibinfo {author} {\bibfnamefont {G.}~\bibnamefont {Dwari}}, \bibinfo {author} {\bibfnamefont {B.~B.}\ \bibnamefont {Maity}}, \bibinfo {author} {\bibfnamefont {V.}~\bibnamefont {Saini}}, \bibinfo {author} {\bibfnamefont {A.}~\bibnamefont {Thamizhavel}},\ and\ \bibinfo {author} {\bibfnamefont {R.}~\bibnamefont {Mondal}},\ }\bibfield  {title} {\bibinfo {title} {Shubnikov-de {Haas} and de {Haas}-van {Alphen} oscillations in {Czochralski} grown {CoSi} single crystal},\ }\href {https://doi.org/10.1088/1361-648X/ac8960} {\bibfield  {journal} {\bibinfo  {journal} {J. Phys.: Condens. Matter}\ }\textbf {\bibinfo {volume} {34}},\ \bibinfo {pages} {425702} (\bibinfo {year} {2022})},\ \bibinfo {note} {publisher: IOP Publishing}\BibitemShut {NoStop}%
\bibitem [{\citenamefont {Petrova}\ \emph {et~al.}(2023)\citenamefont {Petrova}, \citenamefont {Sobolevskii},\ and\ \citenamefont {Stishov}}]{petrova_magnetoresistance_2023}%
  \BibitemOpen
  \bibfield  {author} {\bibinfo {author} {\bibfnamefont {A.~E.}\ \bibnamefont {Petrova}}, \bibinfo {author} {\bibfnamefont {O.~A.}\ \bibnamefont {Sobolevskii}},\ and\ \bibinfo {author} {\bibfnamefont {S.~M.}\ \bibnamefont {Stishov}},\ }\bibfield  {title} {\bibinfo {title} {Magnetoresistance and {Kohler} rule in the topological chiral semimetals {CoSi}},\ }\href {https://doi.org/10.1103/PhysRevB.107.085136} {\bibfield  {journal} {\bibinfo  {journal} {Phys. Rev. B}\ }\textbf {\bibinfo {volume} {107}},\ \bibinfo {pages} {085136} (\bibinfo {year} {2023})},\ \bibinfo {note} {arXiv:2209.02036 [cond-mat]}\BibitemShut {NoStop}%
\bibitem [{\citenamefont {Schnatmann}\ \emph {et~al.}(2020)\citenamefont {Schnatmann}, \citenamefont {Geishendorf}, \citenamefont {Lammel}, \citenamefont {Damm}, \citenamefont {Novikov}, \citenamefont {Thomas}, \citenamefont {Burkov}, \citenamefont {Reith}, \citenamefont {Nielsch},\ and\ \citenamefont {Schierning}}]{schnatmann_signatures_2020}%
  \BibitemOpen
  \bibfield  {author} {\bibinfo {author} {\bibfnamefont {L.}~\bibnamefont {Schnatmann}}, \bibinfo {author} {\bibfnamefont {K.}~\bibnamefont {Geishendorf}}, \bibinfo {author} {\bibfnamefont {M.}~\bibnamefont {Lammel}}, \bibinfo {author} {\bibfnamefont {C.}~\bibnamefont {Damm}}, \bibinfo {author} {\bibfnamefont {S.}~\bibnamefont {Novikov}}, \bibinfo {author} {\bibfnamefont {A.}~\bibnamefont {Thomas}}, \bibinfo {author} {\bibfnamefont {A.}~\bibnamefont {Burkov}}, \bibinfo {author} {\bibfnamefont {H.}~\bibnamefont {Reith}}, \bibinfo {author} {\bibfnamefont {K.}~\bibnamefont {Nielsch}},\ and\ \bibinfo {author} {\bibfnamefont {G.}~\bibnamefont {Schierning}},\ }\bibfield  {title} {\bibinfo {title} {Signatures of a {Charge} {Density} {Wave} {Phase} and the {Chiral} {Anomaly} in the {Fermionic} {Material} {Cobalt} {Monosilicide} {CoSi}},\ }\href {https://doi.org/10.1002/aelm.201900857} {\bibfield  {journal} {\bibinfo  {journal} {Advanced Electronic Materials}\ }\textbf {\bibinfo {volume} {6}},\ \bibinfo {pages}
  {1900857} (\bibinfo {year} {2020})},\ \bibinfo {note} {\_eprint: https://onlinelibrary.wiley.com/doi/pdf/10.1002/aelm.201900857}\BibitemShut {NoStop}%
\bibitem [{\citenamefont {Zhang}\ \emph {et~al.}(2016)\citenamefont {Zhang}, \citenamefont {Xu}, \citenamefont {Belopolski}, \citenamefont {Yuan}, \citenamefont {Lin}, \citenamefont {Tong}, \citenamefont {Bian}, \citenamefont {Alidoust}, \citenamefont {Lee}, \citenamefont {Huang}, \citenamefont {Chang}, \citenamefont {Chang}, \citenamefont {Hsu}, \citenamefont {Jeng}, \citenamefont {Neupane}, \citenamefont {Sanchez}, \citenamefont {Zheng}, \citenamefont {Wang}, \citenamefont {Lin}, \citenamefont {Zhang}, \citenamefont {Lu}, \citenamefont {Shen}, \citenamefont {Neupert}, \citenamefont {Zahid~Hasan},\ and\ \citenamefont {Jia}}]{zhang_signatures_2016}%
  \BibitemOpen
  \bibfield  {author} {\bibinfo {author} {\bibfnamefont {C.-L.}\ \bibnamefont {Zhang}}, \bibinfo {author} {\bibfnamefont {S.-Y.}\ \bibnamefont {Xu}}, \bibinfo {author} {\bibfnamefont {I.}~\bibnamefont {Belopolski}}, \bibinfo {author} {\bibfnamefont {Z.}~\bibnamefont {Yuan}}, \bibinfo {author} {\bibfnamefont {Z.}~\bibnamefont {Lin}}, \bibinfo {author} {\bibfnamefont {B.}~\bibnamefont {Tong}}, \bibinfo {author} {\bibfnamefont {G.}~\bibnamefont {Bian}}, \bibinfo {author} {\bibfnamefont {N.}~\bibnamefont {Alidoust}}, \bibinfo {author} {\bibfnamefont {C.-C.}\ \bibnamefont {Lee}}, \bibinfo {author} {\bibfnamefont {S.-M.}\ \bibnamefont {Huang}}, \bibinfo {author} {\bibfnamefont {T.-R.}\ \bibnamefont {Chang}}, \bibinfo {author} {\bibfnamefont {G.}~\bibnamefont {Chang}}, \bibinfo {author} {\bibfnamefont {C.-H.}\ \bibnamefont {Hsu}}, \bibinfo {author} {\bibfnamefont {H.-T.}\ \bibnamefont {Jeng}}, \bibinfo {author} {\bibfnamefont {M.}~\bibnamefont {Neupane}}, \bibinfo {author} {\bibfnamefont {D.~S.}\ \bibnamefont
  {Sanchez}}, \bibinfo {author} {\bibfnamefont {H.}~\bibnamefont {Zheng}}, \bibinfo {author} {\bibfnamefont {J.}~\bibnamefont {Wang}}, \bibinfo {author} {\bibfnamefont {H.}~\bibnamefont {Lin}}, \bibinfo {author} {\bibfnamefont {C.}~\bibnamefont {Zhang}}, \bibinfo {author} {\bibfnamefont {H.-Z.}\ \bibnamefont {Lu}}, \bibinfo {author} {\bibfnamefont {S.-Q.}\ \bibnamefont {Shen}}, \bibinfo {author} {\bibfnamefont {T.}~\bibnamefont {Neupert}}, \bibinfo {author} {\bibfnamefont {M.}~\bibnamefont {Zahid~Hasan}},\ and\ \bibinfo {author} {\bibfnamefont {S.}~\bibnamefont {Jia}},\ }\bibfield  {title} {\bibinfo {title} {Signatures of the {Adler}–{Bell}–{Jackiw} chiral anomaly in a {Weyl} fermion semimetal},\ }\href {https://doi.org/10.1038/ncomms10735} {\bibfield  {journal} {\bibinfo  {journal} {Nat Commun}\ }\textbf {\bibinfo {volume} {7}},\ \bibinfo {pages} {10735} (\bibinfo {year} {2016})},\ \bibinfo {note} {number: 1 Publisher: Nature Publishing Group}\BibitemShut {NoStop}%
\bibitem [{\citenamefont {Huang}\ \emph {et~al.}(2015)\citenamefont {Huang}, \citenamefont {Zhao}, \citenamefont {Long}, \citenamefont {Wang}, \citenamefont {Chen}, \citenamefont {Yang}, \citenamefont {Liang}, \citenamefont {Xue}, \citenamefont {Weng}, \citenamefont {Fang}, \citenamefont {Dai},\ and\ \citenamefont {Chen}}]{huang_observation_2015}%
  \BibitemOpen
  \bibfield  {author} {\bibinfo {author} {\bibfnamefont {X.}~\bibnamefont {Huang}}, \bibinfo {author} {\bibfnamefont {L.}~\bibnamefont {Zhao}}, \bibinfo {author} {\bibfnamefont {Y.}~\bibnamefont {Long}}, \bibinfo {author} {\bibfnamefont {P.}~\bibnamefont {Wang}}, \bibinfo {author} {\bibfnamefont {D.}~\bibnamefont {Chen}}, \bibinfo {author} {\bibfnamefont {Z.}~\bibnamefont {Yang}}, \bibinfo {author} {\bibfnamefont {H.}~\bibnamefont {Liang}}, \bibinfo {author} {\bibfnamefont {M.}~\bibnamefont {Xue}}, \bibinfo {author} {\bibfnamefont {H.}~\bibnamefont {Weng}}, \bibinfo {author} {\bibfnamefont {Z.}~\bibnamefont {Fang}}, \bibinfo {author} {\bibfnamefont {X.}~\bibnamefont {Dai}},\ and\ \bibinfo {author} {\bibfnamefont {G.}~\bibnamefont {Chen}},\ }\bibfield  {title} {\bibinfo {title} {Observation of the {Chiral}-{Anomaly}-{Induced} {Negative} {Magnetoresistance} in {3D} {Weyl} {Semimetal} {TaAs}},\ }\href {https://doi.org/10.1103/PhysRevX.5.031023} {\bibfield  {journal} {\bibinfo  {journal} {Phys. Rev. X}\ }\textbf
  {\bibinfo {volume} {5}},\ \bibinfo {pages} {031023} (\bibinfo {year} {2015})},\ \bibinfo {note} {publisher: American Physical Society}\BibitemShut {NoStop}%
\bibitem [{\citenamefont {Wang}\ \emph {et~al.}(2016)\citenamefont {Wang}, \citenamefont {Liu}, \citenamefont {Liu}, \citenamefont {Pan}, \citenamefont {Zhang}, \citenamefont {Zeng}, \citenamefont {Fu}, \citenamefont {Wang}, \citenamefont {Xu}, \citenamefont {Huang}, \citenamefont {Wang}, \citenamefont {Lu}, \citenamefont {Xing}, \citenamefont {Wang}, \citenamefont {Wan},\ and\ \citenamefont {Miao}}]{wang_gate-tunable_2016}%
  \BibitemOpen
  \bibfield  {author} {\bibinfo {author} {\bibfnamefont {Y.}~\bibnamefont {Wang}}, \bibinfo {author} {\bibfnamefont {E.}~\bibnamefont {Liu}}, \bibinfo {author} {\bibfnamefont {H.}~\bibnamefont {Liu}}, \bibinfo {author} {\bibfnamefont {Y.}~\bibnamefont {Pan}}, \bibinfo {author} {\bibfnamefont {L.}~\bibnamefont {Zhang}}, \bibinfo {author} {\bibfnamefont {J.}~\bibnamefont {Zeng}}, \bibinfo {author} {\bibfnamefont {Y.}~\bibnamefont {Fu}}, \bibinfo {author} {\bibfnamefont {M.}~\bibnamefont {Wang}}, \bibinfo {author} {\bibfnamefont {K.}~\bibnamefont {Xu}}, \bibinfo {author} {\bibfnamefont {Z.}~\bibnamefont {Huang}}, \bibinfo {author} {\bibfnamefont {Z.}~\bibnamefont {Wang}}, \bibinfo {author} {\bibfnamefont {H.-Z.}\ \bibnamefont {Lu}}, \bibinfo {author} {\bibfnamefont {D.}~\bibnamefont {Xing}}, \bibinfo {author} {\bibfnamefont {B.}~\bibnamefont {Wang}}, \bibinfo {author} {\bibfnamefont {X.}~\bibnamefont {Wan}},\ and\ \bibinfo {author} {\bibfnamefont {F.}~\bibnamefont {Miao}},\ }\bibfield  {title} {\bibinfo {title}
  {Gate-tunable negative longitudinal magnetoresistance in the predicted type-{II} {Weyl} semimetal {WTe2}},\ }\href {https://doi.org/10.1038/ncomms13142} {\bibfield  {journal} {\bibinfo  {journal} {Nat Commun}\ }\textbf {\bibinfo {volume} {7}},\ \bibinfo {pages} {13142} (\bibinfo {year} {2016})},\ \bibinfo {note} {number: 1 Publisher: Nature Publishing Group}\BibitemShut {NoStop}%
\bibitem [{\citenamefont {Guo}\ \emph {et~al.}(2018)\citenamefont {Guo}, \citenamefont {Wu}, \citenamefont {Wu}, \citenamefont {Smidman}, \citenamefont {Cao}, \citenamefont {Bostwick}, \citenamefont {Jozwiak}, \citenamefont {Rotenberg}, \citenamefont {Liu}, \citenamefont {Steglich},\ and\ \citenamefont {Yuan}}]{guo_evidence_2018}%
  \BibitemOpen
  \bibfield  {author} {\bibinfo {author} {\bibfnamefont {C.}~\bibnamefont {Guo}}, \bibinfo {author} {\bibfnamefont {F.}~\bibnamefont {Wu}}, \bibinfo {author} {\bibfnamefont {Z.}~\bibnamefont {Wu}}, \bibinfo {author} {\bibfnamefont {M.}~\bibnamefont {Smidman}}, \bibinfo {author} {\bibfnamefont {C.}~\bibnamefont {Cao}}, \bibinfo {author} {\bibfnamefont {A.}~\bibnamefont {Bostwick}}, \bibinfo {author} {\bibfnamefont {C.}~\bibnamefont {Jozwiak}}, \bibinfo {author} {\bibfnamefont {E.}~\bibnamefont {Rotenberg}}, \bibinfo {author} {\bibfnamefont {Y.}~\bibnamefont {Liu}}, \bibinfo {author} {\bibfnamefont {F.}~\bibnamefont {Steglich}},\ and\ \bibinfo {author} {\bibfnamefont {H.}~\bibnamefont {Yuan}},\ }\bibfield  {title} {\bibinfo {title} {Evidence for {Weyl} fermions in a canonical heavy-fermion semimetal {YbPtBi}},\ }\bibfield  {journal} {\bibinfo  {journal} {Nature Communications}\ }\textbf {\bibinfo {volume} {9}},\ \href {https://doi.org/10.1038/s41467-018-06782-1} {10.1038/s41467-018-06782-1} (\bibinfo {year}
  {2018})\BibitemShut {NoStop}%
\bibitem [{\citenamefont {Liang}\ \emph {et~al.}(2018{\natexlab{b}})\citenamefont {Liang}, \citenamefont {Lin}, \citenamefont {Gibson}, \citenamefont {Kushwaha}, \citenamefont {Liu}, \citenamefont {Wang}, \citenamefont {Xiong}, \citenamefont {Sobota}, \citenamefont {Hashimoto}, \citenamefont {Kirchmann}, \citenamefont {Shen}, \citenamefont {Cava},\ and\ \citenamefont {Ong}}]{liang_anomalous_2018}%
  \BibitemOpen
  \bibfield  {author} {\bibinfo {author} {\bibfnamefont {T.}~\bibnamefont {Liang}}, \bibinfo {author} {\bibfnamefont {J.}~\bibnamefont {Lin}}, \bibinfo {author} {\bibfnamefont {Q.}~\bibnamefont {Gibson}}, \bibinfo {author} {\bibfnamefont {S.}~\bibnamefont {Kushwaha}}, \bibinfo {author} {\bibfnamefont {M.}~\bibnamefont {Liu}}, \bibinfo {author} {\bibfnamefont {W.}~\bibnamefont {Wang}}, \bibinfo {author} {\bibfnamefont {H.}~\bibnamefont {Xiong}}, \bibinfo {author} {\bibfnamefont {J.~A.}\ \bibnamefont {Sobota}}, \bibinfo {author} {\bibfnamefont {M.}~\bibnamefont {Hashimoto}}, \bibinfo {author} {\bibfnamefont {P.~S.}\ \bibnamefont {Kirchmann}}, \bibinfo {author} {\bibfnamefont {Z.-X.}\ \bibnamefont {Shen}}, \bibinfo {author} {\bibfnamefont {R.~J.}\ \bibnamefont {Cava}},\ and\ \bibinfo {author} {\bibfnamefont {N.~P.}\ \bibnamefont {Ong}},\ }\bibfield  {title} {\bibinfo {title} {Anomalous {Hall} effect in {ZrTe5}},\ }\href {https://doi.org/10.1038/s41567-018-0078-z} {\bibfield  {journal} {\bibinfo  {journal}
  {Nature Phys}\ }\textbf {\bibinfo {volume} {14}},\ \bibinfo {pages} {451} (\bibinfo {year} {2018}{\natexlab{b}})},\ \bibinfo {note} {number: 5 Publisher: Nature Publishing Group}\BibitemShut {NoStop}%
\bibitem [{\citenamefont {Pippard}(1989)}]{pippard_magnetoresistance_nodate}%
  \BibitemOpen
  \bibfield  {author} {\bibinfo {author} {\bibfnamefont {A.~B.}\ \bibnamefont {Pippard}},\ }\href@noop {} {\emph {\bibinfo {title} {Magnetoresistance in {Metals}}}}\ (\bibinfo  {publisher} {Cambridge University Press},\ \bibinfo {year} {1989})\BibitemShut {NoStop}%
\bibitem [{\citenamefont {Naumann}\ \emph {et~al.}(2020)\citenamefont {Naumann}, \citenamefont {Arnold}, \citenamefont {Bachmann}, \citenamefont {Modic}, \citenamefont {Moll}, \citenamefont {Süß}, \citenamefont {Schmidt},\ and\ \citenamefont {Hassinger}}]{naumann_orbital_2020}%
  \BibitemOpen
  \bibfield  {author} {\bibinfo {author} {\bibfnamefont {M.}~\bibnamefont {Naumann}}, \bibinfo {author} {\bibfnamefont {F.}~\bibnamefont {Arnold}}, \bibinfo {author} {\bibfnamefont {M.~D.}\ \bibnamefont {Bachmann}}, \bibinfo {author} {\bibfnamefont {K.~A.}\ \bibnamefont {Modic}}, \bibinfo {author} {\bibfnamefont {P.~J.~W.}\ \bibnamefont {Moll}}, \bibinfo {author} {\bibfnamefont {V.}~\bibnamefont {Süß}}, \bibinfo {author} {\bibfnamefont {M.}~\bibnamefont {Schmidt}},\ and\ \bibinfo {author} {\bibfnamefont {E.}~\bibnamefont {Hassinger}},\ }\bibfield  {title} {\bibinfo {title} {Orbital effect and weak localization in the longitudinal magnetoresistance of {Weyl} semimetals {NbP}, {NbAs}, {TaP}, and {TaAs}},\ }\href {https://doi.org/10.1103/PhysRevMaterials.4.034201} {\bibfield  {journal} {\bibinfo  {journal} {Phys. Rev. Mater.}\ }\textbf {\bibinfo {volume} {4}},\ \bibinfo {pages} {034201} (\bibinfo {year} {2020})},\ \bibinfo {note} {publisher: American Physical Society}\BibitemShut {NoStop}%
\bibitem [{\citenamefont {Breunig}\ \emph {et~al.}(2017)\citenamefont {Breunig}, \citenamefont {Wang}, \citenamefont {Taskin}, \citenamefont {Lux}, \citenamefont {Rosch},\ and\ \citenamefont {Ando}}]{breunig_gigantic_2017}%
  \BibitemOpen
  \bibfield  {author} {\bibinfo {author} {\bibfnamefont {O.}~\bibnamefont {Breunig}}, \bibinfo {author} {\bibfnamefont {Z.}~\bibnamefont {Wang}}, \bibinfo {author} {\bibfnamefont {A.~A.}\ \bibnamefont {Taskin}}, \bibinfo {author} {\bibfnamefont {J.}~\bibnamefont {Lux}}, \bibinfo {author} {\bibfnamefont {A.}~\bibnamefont {Rosch}},\ and\ \bibinfo {author} {\bibfnamefont {Y.}~\bibnamefont {Ando}},\ }\bibfield  {title} {\bibinfo {title} {Gigantic negative magnetoresistance in the bulk of a disordered topological insulator},\ }\href {https://doi.org/10.1038/ncomms15545} {\bibfield  {journal} {\bibinfo  {journal} {Nat Commun}\ }\textbf {\bibinfo {volume} {8}},\ \bibinfo {pages} {15545} (\bibinfo {year} {2017})},\ \bibinfo {note} {number: 1 Publisher: Nature Publishing Group}\BibitemShut {NoStop}%
\bibitem [{\citenamefont {Schumann}\ \emph {et~al.}(2017)\citenamefont {Schumann}, \citenamefont {Goyal}, \citenamefont {Kealhofer},\ and\ \citenamefont {Stemmer}}]{schumann_negative_2017}%
  \BibitemOpen
  \bibfield  {author} {\bibinfo {author} {\bibfnamefont {T.}~\bibnamefont {Schumann}}, \bibinfo {author} {\bibfnamefont {M.}~\bibnamefont {Goyal}}, \bibinfo {author} {\bibfnamefont {D.~A.}\ \bibnamefont {Kealhofer}},\ and\ \bibinfo {author} {\bibfnamefont {S.}~\bibnamefont {Stemmer}},\ }\bibfield  {title} {\bibinfo {title} {Negative magnetoresistance due to conductivity fluctuations in films of the topological semimetal \${\textbackslash}mathrm\{{C}\}\{{\textbackslash}mathrm\{d\}\}\_\{3\}{\textbackslash}mathrm\{{A}\}\{{\textbackslash}mathrm\{s\}\}\_\{2\}\$},\ }\href {https://doi.org/10.1103/PhysRevB.95.241113} {\bibfield  {journal} {\bibinfo  {journal} {Phys. Rev. B}\ }\textbf {\bibinfo {volume} {95}},\ \bibinfo {pages} {241113} (\bibinfo {year} {2017})},\ \bibinfo {note} {publisher: American Physical Society}\BibitemShut {NoStop}%
\bibitem [{\citenamefont {Molinari}\ \emph {et~al.}(2023)\citenamefont {Molinari}, \citenamefont {Balduini}, \citenamefont {Rocchino}, \citenamefont {Wawrzyńczak}, \citenamefont {Sousa}, \citenamefont {Bui}, \citenamefont {Lavoie}, \citenamefont {Stanic}, \citenamefont {Jordan-Sweet}, \citenamefont {Hopstaken}, \citenamefont {Tchoumakov}, \citenamefont {Franca}, \citenamefont {Gooth}, \citenamefont {Fratini}, \citenamefont {Grushin}, \citenamefont {Zota}, \citenamefont {Gotsmann},\ and\ \citenamefont {Schmid}}]{molinari_disorder-induced_2023}%
  \BibitemOpen
  \bibfield  {author} {\bibinfo {author} {\bibfnamefont {A.}~\bibnamefont {Molinari}}, \bibinfo {author} {\bibfnamefont {F.}~\bibnamefont {Balduini}}, \bibinfo {author} {\bibfnamefont {L.}~\bibnamefont {Rocchino}}, \bibinfo {author} {\bibfnamefont {R.}~\bibnamefont {Wawrzyńczak}}, \bibinfo {author} {\bibfnamefont {M.}~\bibnamefont {Sousa}}, \bibinfo {author} {\bibfnamefont {H.}~\bibnamefont {Bui}}, \bibinfo {author} {\bibfnamefont {C.}~\bibnamefont {Lavoie}}, \bibinfo {author} {\bibfnamefont {V.}~\bibnamefont {Stanic}}, \bibinfo {author} {\bibfnamefont {J.}~\bibnamefont {Jordan-Sweet}}, \bibinfo {author} {\bibfnamefont {M.}~\bibnamefont {Hopstaken}}, \bibinfo {author} {\bibfnamefont {S.}~\bibnamefont {Tchoumakov}}, \bibinfo {author} {\bibfnamefont {S.}~\bibnamefont {Franca}}, \bibinfo {author} {\bibfnamefont {J.}~\bibnamefont {Gooth}}, \bibinfo {author} {\bibfnamefont {S.}~\bibnamefont {Fratini}}, \bibinfo {author} {\bibfnamefont {A.~G.}\ \bibnamefont {Grushin}}, \bibinfo {author} {\bibfnamefont
  {C.}~\bibnamefont {Zota}}, \bibinfo {author} {\bibfnamefont {B.}~\bibnamefont {Gotsmann}},\ and\ \bibinfo {author} {\bibfnamefont {H.}~\bibnamefont {Schmid}},\ }\bibfield  {title} {\bibinfo {title} {Disorder-{Induced} {Magnetotransport} {Anomalies} in {Amorphous} and {Textured} {Co1}–{xSix} {Semimetal} {Thin} {Films}},\ }\href {https://doi.org/10.1021/acsaelm.3c00095} {\bibfield  {journal} {\bibinfo  {journal} {ACS Appl. Electron. Mater.}\ }\textbf {\bibinfo {volume} {5}},\ \bibinfo {pages} {2624} (\bibinfo {year} {2023})},\ \bibinfo {note} {publisher: American Chemical Society}\BibitemShut {NoStop}%
\bibitem [{\citenamefont {Behrends}\ and\ \citenamefont {Bardarson}(2017)}]{Behrends2017}%
  \BibitemOpen
  \bibfield  {author} {\bibinfo {author} {\bibfnamefont {J.}~\bibnamefont {Behrends}}\ and\ \bibinfo {author} {\bibfnamefont {J.~H.}\ \bibnamefont {Bardarson}},\ }\bibfield  {title} {\bibinfo {title} {Strongly angle-dependent magnetoresistance in weyl semimetals with long-range disorder},\ }\href {https://doi.org/10.1103/PhysRevB.96.060201} {\bibfield  {journal} {\bibinfo  {journal} {Phys. Rev. B}\ }\textbf {\bibinfo {volume} {96}},\ \bibinfo {pages} {060201} (\bibinfo {year} {2017})}\BibitemShut {NoStop}%
\bibitem [{\citenamefont {Deng}\ \emph {et~al.}(2020)\citenamefont {Deng}, \citenamefont {Ba}, \citenamefont {Ma}, \citenamefont {Luo}, \citenamefont {Wang}, \citenamefont {Sheng},\ and\ \citenamefont {Xing}}]{Deng2020}%
  \BibitemOpen
  \bibfield  {author} {\bibinfo {author} {\bibfnamefont {M.-X.}\ \bibnamefont {Deng}}, \bibinfo {author} {\bibfnamefont {J.-Y.}\ \bibnamefont {Ba}}, \bibinfo {author} {\bibfnamefont {R.}~\bibnamefont {Ma}}, \bibinfo {author} {\bibfnamefont {W.}~\bibnamefont {Luo}}, \bibinfo {author} {\bibfnamefont {R.-Q.}\ \bibnamefont {Wang}}, \bibinfo {author} {\bibfnamefont {L.}~\bibnamefont {Sheng}},\ and\ \bibinfo {author} {\bibfnamefont {D.~Y.}\ \bibnamefont {Xing}},\ }\bibfield  {title} {\bibinfo {title} {Chiral-anomaly-induced angular narrowing of the positive longitudinal magnetoconductivity in weyl semimetals},\ }\href {https://doi.org/10.1103/PhysRevResearch.2.033346} {\bibfield  {journal} {\bibinfo  {journal} {Phys. Rev. Res.}\ }\textbf {\bibinfo {volume} {2}},\ \bibinfo {pages} {033346} (\bibinfo {year} {2020})}\BibitemShut {NoStop}%
\bibitem [{\citenamefont {Das}\ and\ \citenamefont {Agarwal}(2021)}]{Agarwal21}%
  \BibitemOpen
  \bibfield  {author} {\bibinfo {author} {\bibfnamefont {K.}~\bibnamefont {Das}}\ and\ \bibinfo {author} {\bibfnamefont {A.}~\bibnamefont {Agarwal}},\ }\bibfield  {title} {\bibinfo {title} {Intrinsic hall conductivities induced by the orbital magnetic moment},\ }\href {https://doi.org/10.1103/PhysRevB.103.125432} {\bibfield  {journal} {\bibinfo  {journal} {Phys. Rev. B}\ }\textbf {\bibinfo {volume} {103}},\ \bibinfo {pages} {125432} (\bibinfo {year} {2021})}\BibitemShut {NoStop}%
\bibitem [{\citenamefont {Nielsen}\ and\ \citenamefont {Ninomiya}(1981{\natexlab{a}})}]{NielNino81a}%
  \BibitemOpen
  \bibfield  {author} {\bibinfo {author} {\bibfnamefont {H.~B.}\ \bibnamefont {Nielsen}}\ and\ \bibinfo {author} {\bibfnamefont {M.}~\bibnamefont {Ninomiya}},\ }\bibfield  {title} {\bibinfo {title} {{Absence of neutrinos on a lattice: (I). Proof by homotopy theory}},\ }\href {https://doi.org/10.1016/0550-3213(81)90361-8} {\bibfield  {journal} {\bibinfo  {journal} {Nuclear Physics B}\ }\textbf {\bibinfo {volume} {185}},\ \bibinfo {pages} {20 } (\bibinfo {year} {1981}{\natexlab{a}})}\BibitemShut {NoStop}%
\bibitem [{\citenamefont {Nielsen}\ and\ \citenamefont {Ninomiya}(1981{\natexlab{b}})}]{NielNino81b}%
  \BibitemOpen
  \bibfield  {author} {\bibinfo {author} {\bibfnamefont {H.~B.}\ \bibnamefont {Nielsen}}\ and\ \bibinfo {author} {\bibfnamefont {M.}~\bibnamefont {Ninomiya}},\ }\bibfield  {title} {\bibinfo {title} {{Absence of neutrinos on a lattice: (II). Intuitive topological proof}},\ }\href {https://doi.org/10.1016/0550-3213(81)90524-1} {\bibfield  {journal} {\bibinfo  {journal} {Nuclear Physics B}\ }\textbf {\bibinfo {volume} {193}},\ \bibinfo {pages} {173 } (\bibinfo {year} {1981}{\natexlab{b}})}\BibitemShut {NoStop}%
\bibitem [{\citenamefont {Lundgren}\ \emph {et~al.}(2014)\citenamefont {Lundgren}, \citenamefont {Laurell},\ and\ \citenamefont {Fiete}}]{Lundgren2014}%
  \BibitemOpen
  \bibfield  {author} {\bibinfo {author} {\bibfnamefont {R.}~\bibnamefont {Lundgren}}, \bibinfo {author} {\bibfnamefont {P.}~\bibnamefont {Laurell}},\ and\ \bibinfo {author} {\bibfnamefont {G.~A.}\ \bibnamefont {Fiete}},\ }\bibfield  {title} {\bibinfo {title} {Thermoelectric properties of weyl and dirac semimetals},\ }\href {https://doi.org/10.1103/PhysRevB.90.165115} {\bibfield  {journal} {\bibinfo  {journal} {Phys. Rev. B}\ }\textbf {\bibinfo {volume} {90}},\ \bibinfo {pages} {165115} (\bibinfo {year} {2014})}\BibitemShut {NoStop}%
\bibitem [{\citenamefont {Kim}\ \emph {et~al.}(2014)\citenamefont {Kim}, \citenamefont {Kim},\ and\ \citenamefont {Sasaki}}]{KimKim2014}%
  \BibitemOpen
  \bibfield  {author} {\bibinfo {author} {\bibfnamefont {K.-S.}\ \bibnamefont {Kim}}, \bibinfo {author} {\bibfnamefont {H.-J.}\ \bibnamefont {Kim}},\ and\ \bibinfo {author} {\bibfnamefont {M.}~\bibnamefont {Sasaki}},\ }\bibfield  {title} {\bibinfo {title} {Boltzmann equation approach to anomalous transport in a weyl metal},\ }\href {https://doi.org/10.1103/PhysRevB.89.195137} {\bibfield  {journal} {\bibinfo  {journal} {Phys. Rev. B}\ }\textbf {\bibinfo {volume} {89}},\ \bibinfo {pages} {195137} (\bibinfo {year} {2014})}\BibitemShut {NoStop}%
\bibitem [{\citenamefont {Ma}\ and\ \citenamefont {Pesin}(2015)}]{MaPesin2015}%
  \BibitemOpen
  \bibfield  {author} {\bibinfo {author} {\bibfnamefont {J.}~\bibnamefont {Ma}}\ and\ \bibinfo {author} {\bibfnamefont {D.~A.}\ \bibnamefont {Pesin}},\ }\bibfield  {title} {\bibinfo {title} {Chiral magnetic effect and natural optical activity in metals with or without weyl points},\ }\href {https://doi.org/10.1103/PhysRevB.92.235205} {\bibfield  {journal} {\bibinfo  {journal} {Phys. Rev. B}\ }\textbf {\bibinfo {volume} {92}},\ \bibinfo {pages} {235205} (\bibinfo {year} {2015})}\BibitemShut {NoStop}%
\bibitem [{\citenamefont {Imran}\ and\ \citenamefont {Hershfield}(2018)}]{Imran2018}%
  \BibitemOpen
  \bibfield  {author} {\bibinfo {author} {\bibfnamefont {M.}~\bibnamefont {Imran}}\ and\ \bibinfo {author} {\bibfnamefont {S.}~\bibnamefont {Hershfield}},\ }\bibfield  {title} {\bibinfo {title} {Berry curvature force and lorentz force comparison in the magnetotransport of weyl semimetals},\ }\href {https://doi.org/10.1103/PhysRevB.98.205139} {\bibfield  {journal} {\bibinfo  {journal} {Phys. Rev. B}\ }\textbf {\bibinfo {volume} {98}},\ \bibinfo {pages} {205139} (\bibinfo {year} {2018})}\BibitemShut {NoStop}%
\bibitem [{\citenamefont {Deng}\ \emph {et~al.}(2019)\citenamefont {Deng}, \citenamefont {Duan}, \citenamefont {Luo}, \citenamefont {Deng}, \citenamefont {Wang},\ and\ \citenamefont {Sheng}}]{Deng2019}%
  \BibitemOpen
  \bibfield  {author} {\bibinfo {author} {\bibfnamefont {M.-X.}\ \bibnamefont {Deng}}, \bibinfo {author} {\bibfnamefont {H.-J.}\ \bibnamefont {Duan}}, \bibinfo {author} {\bibfnamefont {W.}~\bibnamefont {Luo}}, \bibinfo {author} {\bibfnamefont {W.~Y.}\ \bibnamefont {Deng}}, \bibinfo {author} {\bibfnamefont {R.-Q.}\ \bibnamefont {Wang}},\ and\ \bibinfo {author} {\bibfnamefont {L.}~\bibnamefont {Sheng}},\ }\bibfield  {title} {\bibinfo {title} {Quantum oscillation modulated angular dependence of the positive longitudinal magnetoconductivity and planar hall effect in weyl semimetals},\ }\href {https://doi.org/10.1103/PhysRevB.99.165146} {\bibfield  {journal} {\bibinfo  {journal} {Phys. Rev. B}\ }\textbf {\bibinfo {volume} {99}},\ \bibinfo {pages} {165146} (\bibinfo {year} {2019})}\BibitemShut {NoStop}%
\bibitem [{\citenamefont {Mandal}\ \emph {et~al.}(2022)\citenamefont {Mandal}, \citenamefont {Das},\ and\ \citenamefont {Agarwal}}]{Mandal2022}%
  \BibitemOpen
  \bibfield  {author} {\bibinfo {author} {\bibfnamefont {D.}~\bibnamefont {Mandal}}, \bibinfo {author} {\bibfnamefont {K.}~\bibnamefont {Das}},\ and\ \bibinfo {author} {\bibfnamefont {A.}~\bibnamefont {Agarwal}},\ }\bibfield  {title} {\bibinfo {title} {{Chiral anomaly and nonlinear magnetotransport in time reversal symmetric Weyl semimetals}},\ }\href {https://doi.org/10.1103/physrevb.106.035423} {\bibfield  {journal} {\bibinfo  {journal} {Physical Review B}\ }\textbf {\bibinfo {volume} {106}},\ \bibinfo {pages} {035423} (\bibinfo {year} {2022})},\ \Eprint {https://arxiv.org/abs/2201.02505} {2201.02505} \BibitemShut {NoStop}%
\bibitem [{\citenamefont {Sundaram}\ and\ \citenamefont {Niu}(1999)}]{Sundaram99}%
  \BibitemOpen
  \bibfield  {author} {\bibinfo {author} {\bibfnamefont {G.}~\bibnamefont {Sundaram}}\ and\ \bibinfo {author} {\bibfnamefont {Q.}~\bibnamefont {Niu}},\ }\bibfield  {title} {{\selectlanguage {English}\bibinfo {title} {{Wave-packet dynamics in slowly perturbed crystals: Gradient corrections and Berry-phase effects}}},\ }\href {https://doi.org/10.1103/physrevb.59.14915} {\bibfield  {journal} {\bibinfo  {journal} {Physical Review B}\ }\textbf {\bibinfo {volume} {59}},\ \bibinfo {pages} {14915 } (\bibinfo {year} {1999})}\BibitemShut {NoStop}%
\bibitem [{\citenamefont {Gao}\ \emph {et~al.}(2014)\citenamefont {Gao}, \citenamefont {Yang},\ and\ \citenamefont {Niu}}]{Gao2014}%
  \BibitemOpen
  \bibfield  {author} {\bibinfo {author} {\bibfnamefont {Y.}~\bibnamefont {Gao}}, \bibinfo {author} {\bibfnamefont {S.~A.}\ \bibnamefont {Yang}},\ and\ \bibinfo {author} {\bibfnamefont {Q.}~\bibnamefont {Niu}},\ }\bibfield  {title} {\bibinfo {title} {Field induced positional shift of bloch electrons and its dynamical implications},\ }\href {https://doi.org/10.1103/PhysRevLett.112.166601} {\bibfield  {journal} {\bibinfo  {journal} {Phys. Rev. Lett.}\ }\textbf {\bibinfo {volume} {112}},\ \bibinfo {pages} {166601} (\bibinfo {year} {2014})}\BibitemShut {NoStop}%
\bibitem [{\citenamefont {Cai}\ \emph {et~al.}(2013)\citenamefont {Cai}, \citenamefont {Yang}, \citenamefont {Li}, \citenamefont {Zhang}, \citenamefont {Shi}, \citenamefont {Yao},\ and\ \citenamefont {Niu}}]{Cai2013}%
  \BibitemOpen
  \bibfield  {author} {\bibinfo {author} {\bibfnamefont {T.}~\bibnamefont {Cai}}, \bibinfo {author} {\bibfnamefont {S.~A.}\ \bibnamefont {Yang}}, \bibinfo {author} {\bibfnamefont {X.}~\bibnamefont {Li}}, \bibinfo {author} {\bibfnamefont {F.}~\bibnamefont {Zhang}}, \bibinfo {author} {\bibfnamefont {J.}~\bibnamefont {Shi}}, \bibinfo {author} {\bibfnamefont {W.}~\bibnamefont {Yao}},\ and\ \bibinfo {author} {\bibfnamefont {Q.}~\bibnamefont {Niu}},\ }\bibfield  {title} {\bibinfo {title} {Magnetic control of the valley degree of freedom of massive dirac fermions with application to transition metal dichalcogenides},\ }\href {https://doi.org/10.1103/PhysRevB.88.115140} {\bibfield  {journal} {\bibinfo  {journal} {Phys. Rev. B}\ }\textbf {\bibinfo {volume} {88}},\ \bibinfo {pages} {115140} (\bibinfo {year} {2013})}\BibitemShut {NoStop}%
\bibitem [{\citenamefont {Suh}\ and\ \citenamefont {Min}(2024)}]{suh2024effect}%
  \BibitemOpen
  \bibfield  {author} {\bibinfo {author} {\bibfnamefont {J.}~\bibnamefont {Suh}}\ and\ \bibinfo {author} {\bibfnamefont {H.}~\bibnamefont {Min}},\ }\bibfield  {title} {\bibinfo {title} {Effect of trivial bands on chiral anomaly-induced longitudinal magnetoconductivity in weyl semimetals},\ }\href@noop {} {\bibfield  {journal} {\bibinfo  {journal} {arXiv:2401.13855}\ } (\bibinfo {year} {2024})},\ \Eprint {https://arxiv.org/abs/2401.13855} {arXiv:2401.13855 [cond-mat.mes-hall]} \BibitemShut {NoStop}%
\bibitem [{\citenamefont {Huber}\ \emph {et~al.}(2023)\citenamefont {Huber}, \citenamefont {Leeb}, \citenamefont {Bauer}, \citenamefont {Benka}, \citenamefont {Knolle}, \citenamefont {Pfleiderer},\ and\ \citenamefont {Wilde}}]{Huber2023}%
  \BibitemOpen
  \bibfield  {author} {\bibinfo {author} {\bibfnamefont {N.}~\bibnamefont {Huber}}, \bibinfo {author} {\bibfnamefont {V.}~\bibnamefont {Leeb}}, \bibinfo {author} {\bibfnamefont {A.}~\bibnamefont {Bauer}}, \bibinfo {author} {\bibfnamefont {G.}~\bibnamefont {Benka}}, \bibinfo {author} {\bibfnamefont {J.}~\bibnamefont {Knolle}}, \bibinfo {author} {\bibfnamefont {C.}~\bibnamefont {Pfleiderer}},\ and\ \bibinfo {author} {\bibfnamefont {M.~A.}\ \bibnamefont {Wilde}},\ }\bibfield  {title} {\bibinfo {title} {{Quantum oscillations of the quasiparticle lifetime in a metal}},\ }\href {https://doi.org/10.1038/s41586-023-06330-y} {\bibfield  {journal} {\bibinfo  {journal} {Nature}\ }\textbf {\bibinfo {volume} {621}},\ \bibinfo {pages} {276} (\bibinfo {year} {2023})},\ \Eprint {https://arxiv.org/abs/2306.09420} {2306.09420} \BibitemShut {NoStop}%
\end{thebibliography}
\end{document}